 \documentstyle[epsf,epsfig,rotate]{aa}
 
\newcommand{\be}{\begin{equation}}
\newcommand{\ee}{\end{equation}}

\newcommand{\ls}{\; \raisebox{-.8ex}{$\buildrel{\textstyle<}\over\sim$}\;}
   
\newcommand{\anrev}{{\it ARA\&A, }}
\newcommand{\apj}{{\it ApJ, }}

\newcommand{\mnr}{{\it MNRAS, }}
\newcommand{\nat}{{\it Nat, }}

\newcommand{\ana}{{\it A\&A, }}

\newcounter{pp3}
\addtocounter{pp3}{3}
   
\begin{document}
\thesaurus{02(02.01.2; 02.13.2; 08.09.2 AA Tau; 08.13.1; 08.16.5)}
     
\title
{Global m=1 modes and migration of 
protoplanetary cores in eccentric protoplanetary  discs}

\author{J.C.B. Papaloizou}
\institute{ Astronomy Unit,
Queen Mary, University of London, Mile End
Rd, London E1 4NS}
	
\offprints{jcbp@maths.qmw.ac.uk}

\date{Received /Accepted}
   
\def\LaTeX{L\kern-.36em\raise.3ex\hbox{a}\kern-.15em
T\kern-.1667em\lower.7ex\hbox{E}\kern-.125emX}
\titlerunning{ Protoplanets embedded in  eccentric discs}
\authorrunning{ J.C.B.   Papaloizou }
\maketitle

\begin{abstract}
We  calculate
global $m=1$ modes with low
pattern speed corresponding
to introducing a finite  eccentricity into a protoplanetary disc.
We consider disc models  which are either isolated
or contain one or two
protoplanets orbiting in an inner cavity.
Global modes
that are strongly coupled to  inner protoplanets
are found to
have disc orbits which tend to have  apsidal lines
antialigned  with respect to those of the inner protoplanets.
Other modes corresponding to free disc modes may be global
over a large range of length scales and accordingly be long lived.
We consider the
motion of a protoplanet in the  earth mass range 
embedded in an eccentric disc
and determine
the equilibrium  orbits  which maintain fixed
apsidal alignment with respect to  the disc gas orbits.
Equilibrium eccentricities are found
to be comparable or possibly  exceed
the disc eccentricity.
We then approximately calculate the tidal interaction with the disc
in order to estimate
the orbital migration rate.   Results are found to deviate
from the case of axisymmetric disc with near circular protoplanet
orbit  once eccentricities of protoplanet and disc  orbits
become comparable to the disc aspect ratio in magnitude.
Aligned protoplanet orbits with very similar
eccentricity to that of the gas disc  are found to
undergo litle  eccentricity change
while undergoing inward migration in general.
However,   for  significantly larger  orbital 
eccentricities,
migration
may be significantly reduced or  even reverse from inwards to outwards.
Thus the existence of global non circular  motions
in discs with radial excursions comparable to the semi-thickness
may have important consequences for the  migration and survival
of protoplanetary  cores in the earth mass range. 
                                                               
\end{abstract}


\section {Introduction}
The recent discovery of a number of extrasolar giant planets orbiting around
nearby solar--type stars ( Mayor \& Queloz, 1995,
Marcy \& Butler 1998, 2000)  reveals that
they  have masses that
are comparable to that of Jupiter,
have orbital semi-major axes in the range

\noindent $0.04 \; {\rm AU} \; \ls \; a \; \ls \; 2.5 \; {\rm au}$, and orbital 
eccentricities in the range $ 0.0 \; \ls \; e \; \ls \; 0.93.$ 

Orbital migration
originally considered as a mechanism operating
in protoplanetary discs by 
Goldreich \& Tremaine (1980) 
has been suggested as an explanation
for the existence of giant planets  close to their  central
star (eg. Lin, Bodenheimer, \& Richardson 1996).
Protoplanetary cores are thought to form at several
astronomical units and then migrate inwards  either before
accumulating a gaseous envelope and while in the earth mass range
(type I migration)
or in the form of a giant planet (type II migration)
(see for example Lin \& Papaloizou 1993, Ward 1997).
In the former case the interaction with the
disc is treated in a linearized approximation
( eg. Goldreich \& Tremaine 1980) while in the latter
nonlinearity is important leading to gap formation
( eg. Bryden et al 1999, Kley 1999).
Both types of migration heve been found to occur
on a timescale more than one order of magnitude
shorter  than that required to form giant planets
or  the  expected lifetime of the disc
and accordingly the survival of embryo protoplanets
is in question ( eg. Ward 1997,  Nelson et al 2000 ).
Accordingly mechanisms that slow or halt migration
such as the entry  into a magnetospheric cavity close
to the star ( Lin, Bodenheimer, \& Richardson 1996)
have been suggested. However, the existence of giant
planets over a range of semi-major axes suggests that a
more general mechanism for halting migration should be sought.

In this paper we shall consider  effects  produced when
the large scale  motion of the disc gas deviates from that of pure
circular motion about the central mass as  could be produced
when the disc becomes eccentric as happens when it supports
a global $m = 1$ mode. Such modes could be produced by disc
protoplanet interactions (eg. Papaloizou, Nelson \& Masset 2001).
However, for standard parameters,
it was found that an instability leading to an eccentric
disc together with an eccentric protoplanet orbit occurred only
for large masses exceeding about fifteen Jupiter masses.

In this context note these calculations were done for laminar model discs with
anomalously 
large viscosity coefficient and results may be somewhat different
for turbulent discs. These simulations as well as resonant
torque calculations appropriate to embedded cores by
Papaloizou \& Larwood (2000) have indicated possible reversal
of orbital migration associated with eccentric orbits
with modest  eccentricity comparable to the disc aspect ratio.

This leads us to study further the dynamics of  protoplanetary
cores, massive enough for tidal interactions with the
disc to be more important than effects
due to gas drag,  in an eccentric protoplanetary disc.
In this paper we consider large scale $m = 1$ modes in gaseous
protoplanetary discs that correspond to making them eccentric.
These modes have a low frequency branch  for which
 the disc gas follows  trajectories
differing from Keplerian eccentric
orbits by small corrections depending on forces
due to disc self-gravity and pressure. These modes can be global
in that they may
vary on a length scale comparable to that of the whole disc
even though it might have  a large dynamic range.
We shall consider a  ratio of outer to inner radius of one hundred.
These modes  are of interest because, even though no definitive
excitation mechanism of general applicability has  yet   been identified,
their large scale implies a long life time comparable to the viscous
time of the disc making them of potential
interest in Astrophysics  ( eg. Ogilvie 2001).  They have also  been
considered as a potential source of angular momentum 
transport by Lee \& Goodman (1999) in a tight winding approximation.
 Furthermore a disc composed of many  stars on near Keplerian orbits 
has been postulated to occur in such objects
as the nucleus of M31 (Tremaine 1995).

Here we consider global $ m  = 1$ modes for various disc models
neglecting viscous processes which are presumed to act over
a longer time scale than that appropriate to the 
phenomena of interest. For global disturbances
in protoplanetary discs of the type we consider, inclusion of self-gravity
is important. Even though the discs are gravitationally stable,
pressure and self-gravity can be equally important on scales
comparable to the current radius, $r,$ when the Toomre stability
parameter $Q \sim r/H,$ with $H$ being the disc semi-thickness.
This condition is satisfied for typical protostellar disc
models ( eg. Papaloizou \& Terquem 1999).  In addition to this,
disc self-gravity has to be considered in the description of the motion
of embedded protoplanets and it generally causes them to move in
eccentric orbits, with eccentricity comparable to or exceeding that
in the disc. We find that circularization due to tidal interaction
with the disc may play only a minor role if the local test particle
precession frequency relative to the disc is large compared to
the circularization rate. We also find that for modest
disc eccentricities comparable to the aspect ratio $H/r,$
disc tidal interactions may differ significantly from those found
in axisymmetric discs. This in turn may have important consequences
for estimates of orbital migration rates
for protoplanetary cores in the  earth mass range.

In section \ref{s1} we give the  Basic equations and the
linearized form  we use to calculate global $m=1$ modes with low
pattern speed corresponding
to introducing a finite disc eccentricity.
We go on to describe the disc models used which may contain
protoplanets orbiting in an inner cavity.
In section \ref{s2}) we present the 
results of normal mode calculations.
We go on to  discuss the
motion of a protoplanet in the earth mass range
in an eccentric disc  in section \ref{s3},        
determining the equilibrium (non precessing)  orbits  which maintain
apsidal alignment with the disc gas orbits.
We then formulate the calculation of the tidal
response of an eccentric disc to a low mass protoplanet in section \ref{s4}
determining the time rate of change of the eccentricity and
orbital migration rate. We find that aligned orbits with very similar
eccentricity to that of the gas disc may suffer no eccentricity change
while undergoing inward migration in general.
However, when the non precessing aligned orbit has a 
significantly higher eccentricity than the disc, as can 
occur for modes with very small pattern speed, orbital migration
may be significantly reduced or reverse from inwards to outwards
for the disc models we consider.
This finding is supported by a local dynamical friction calculation
applicable  when the  protoplanet eccentricity is much larger 
than the disc aspect ratio.
Finally we go on to discuss our results in section \ref{conclusion}.

\section{Basic equations and their linearized form
for large scale global modes with $m=1.$} \label{s1}
We work in a non rotating cylindrical coordinate system  $(r,\varphi)$ 
which may initially be considered to be centred on the primary star.
The basic equations of motion  in a two dimensional approximation
which should be appropriate for  a large scale description
of the disc are taken to be
\begin{equation}  {\partial  v_{r} \over\partial t}+
v_{r}{\partial  v_{r} \over\partial r}
+{v_{\varphi}\over r}{\partial  v_{r} \over\partial \varphi}
-{v_{\varphi}^2 \over r} =
-{1\over \Sigma}{\partial \Pi \over\partial r}-
{\partial \Phi   \over\partial r},
\label{12d} \end{equation}
 
\begin{equation}  {\partial  v_{\varphi} \over\partial t}+
v_{r}{\partial  v_{\varphi} \over\partial r}
+{v_{\varphi}\over r}{\partial  v_{\varphi} \over\partial \varphi}
+{v_{\varphi} v_r\over r} =
-{1\over r\Sigma}{\partial \Pi \over\partial \varphi}-
{1\over r} {\partial \Phi  \over\partial \varphi },
\label{12e} \end{equation}
In addition we have the two dimensional form of the continuity equation
\begin{equation}
{\partial (r\Sigma) \over\partial t}+ {\partial \left(r\Sigma v_r\right)\over \partial r}
+{\partial \left(r\Sigma v_{\varphi}
\right)\over\partial \varphi}=0.\label{12f} \end{equation}
 
Here the velocity is ${\bf v} = ( v_r , v_{\varphi}),$
 $\Pi=\int^{\infty}_{-\infty}Pdz$
represents a vertically integrated pressure
which we assume to be a function of  the surface density $\Sigma$ defined by
\begin{equation}
\Sigma =\int^{\infty}_{-\infty} \rho dz. \end{equation}
The sound speed is then given by  $c = \sqrt{d\Pi /d \Sigma}.$
The gravitational
potential $\Phi= -GM_*/r +\Phi_D + \Phi_{ext},$
with $G$ being the gravitational constant,
has a point mass  contribution  arising from
the
central mass
$M_*,$ a contribution, $\Phi_{D},$ due to the disc
and a contribution, $\Phi_{ext},$ due to external protoplanets.

The unperturbed disk is axisymmetric
with no radial motion such that
the velocity ${\bf v}=(0,r\Omega)$ with $\Omega > 0.$
In equilibrium we then have from equation (\ref{12d})
\begin{equation} 
{\Omega^2 r} =
{1\over \Sigma}{\partial \Pi \over\partial r}+
{\partial \Phi   \over\partial r}
\end{equation}
and the gravitational
potential 
due to the  disc is given by 
\be \Phi_D = -G\int \Sigma K(r,r')r'dr',\label{pot1}\ee
with
\be K_0(r,r') = \int^{2\pi}_0
{1 \over \sqrt{(r^2+r'^2 -2rr'\cos(\varphi))}}
d\varphi.\ee

We  allow for the effects of orbiting external protoplanets
which provide
the external potential $\Phi_{ext}.$
 In this paper we are primarily interested
in phenomena which vary on a time scale
long compared to a local orbital period.
Accordingly we use the time  averaged or secular
protoplanet  perturbing potential which is 
derived in Appendix 1.  For a single protoplanet
of mass $m_p$ and orbital radius $r_p,$ the  contribution
to the external potential is given by
\be \Phi_{ext} = -G { m_p \over 2\pi }
 K_0(r,r_p).\label{potpp1}\ee

In a thin disc  of the kind considered here, the contributions
due to pressure and self-gravity  are comparable and  small  leading to
nearly Keplerian rotation  which has $\Omega \propto r^{-3/2}.$
When  $c \propto r^{-1/2},$
the disc then   has a  nearly constant  putative 
aspect ratio $H/r = c/(r\Omega),$ $H$ being the putative semi thickness.
 
\subsection{Linearization}
Here we are interested in global $m=1$ modes  
with slowly varying pattern when viewed in  the adopted reference frame. 
To study these, we linearize the basic equations
about the equilibrium state
denoting perturbations to
quantities  with a prime. 
 
As usual we assume that the dependence of  all perturbations
(in cylindrical coordinates)  on $\varphi$ and $t$
is through a factor $\exp i\left( m\left (\varphi -\Omega_p t\right) \right).$
For  the slowly varying modes we consider $\Omega_p << \Omega.$
 
The linearized forms of equations (\ref{12d}) ,(\ref{12e})
and (\ref{12f})  are
then
\begin{equation}  im(\Omega-\Omega_p) v'_{r} -
2\Omega v'_{\varphi} =
-{\partial W \over\partial r},
\label{12dp} \end{equation}
 
\begin{equation} im(\Omega-\Omega_p) v'_{\varphi}
+ {\kappa^2\over 2\Omega} v'_{r} =
-{im W\over r},
\label{12ep} \end{equation}
and
\begin{equation}
{im (\Omega-\Omega_p)\Sigma ( W-\phi') \over c^2}=-
{1\over r}{\partial \left(r\Sigma v'_r\right)\over \partial r}
-{im \Sigma v'_{\varphi}
\over r} .\label{12fp}  \end{equation}
 
Here $W = \Pi'/\Sigma +\Phi' = \Sigma'c^2/\Sigma+\Phi',$

\noindent and $\kappa^2 =(2\Omega/r)(d(r^2\Omega)/dr)$
denotes the square of the epicyclic frequency.

For the gravitational potential perturbation we have
$\Phi' =\Phi_D'+\Phi'_{ext}.$
Here we shall  allow
for the contribution
of the secular effects of other sources
such as protoplanets to the modes through the
gravitational potential perturbation $\Phi'_{ext}.$
 
\subsection{ Slowly varying  modes with m=1}
 
We consider linear perturbations of an   
axisymmetric disc  that correspond to normal modes with $m=1$
that make it eccentric. The modes we consider  are
such that $\Omega_p << \Omega$ and the dominant
motion is epicyclic. In this case, to lowest order,
we may neglect the pressure and self-gravity term $W$
as well as $\Omega_p$ and assume $\Omega$
is keplerian, to obtain from equation (\ref{12ep})
\be   i v'_{\varphi}
=- {1\over 2} v'_{r}.\ee
 Using the above and introducing the radial  Lagrangian displacement 
$\xi _r = -i v'_{r}/(\Omega - \Omega_p),$ 
equation (\ref{12fp}) gives
the
surface density perturbation  in the low frequency
limit in the form
\be \Sigma'= -r{d ( \Sigma e(r) )\over dr},  \label{CRP}\ee
with $e(r) \equiv \xi_r/r$ being the disc eccentricity.

The gravitational potential perturbation induced by the disc  is
\be \Phi'_D = -G\int \Sigma' K_1(r,r')r'dr',\label{potpert1}\ee
with
\be K_1(r,r') = \int^{2\pi}_0
{\cos(\varphi) \over \sqrt{(r^2+r'^2 -2rr'\cos(\varphi))}}
d\varphi -{\pi r\over  r'^2},\ee
where the second term corresponds to the indirect term
which results from the acceleration  of the coordinate system
produced by the disc material.

The existence of a mode with $m=1$ in the disc causes
the orbits of the  protoplanets
interior to the disc to become eccentric.
 Noting the different form of indirect term used,
the secular perturbing potential
derived in Appendix 1 then gives 
 the contribution of an orbiting protoplanet
to the external potential perturbation to  be 
\be \Phi'_p(r) = -
{G m_p \xi_r(r_p) \over 2\pi r_p^2}
{\partial  \over \partial r_p}\left(
K_1(r,r_p)
r_p^2 \right).\label{potplt}
\ee
Here the eccentricity of the  
companion orbit is related to the displacement by
$e_p = \xi_r(r_p)/r_p.$
The   total  external potential perturbation is then found by summing over the 
perturbing protoplanets:

\be \Phi'_{ext}(r) = \sum_p \Phi'_p(r) \ee 
 
Eliminating
$v_{\varphi}'$ from equations (\ref{12dp}) and   (\ref{12ep}), 
using  equation (\ref{CRP})  for $\Sigma'$ and expanding to first order
in the small frequencies $\Omega_p$ and $\omega_p =\Omega - \kappa,$
the  fluid element  orbital precession frequency in the absence
of surface density or gravitational potential
perturbations,
one obtains a
normal mode equation relating $e(r), e_p$  and $\Omega_p$
in  the form (see also Papaloizou, Nelson \& Masset, 2001)
\be 2\left(\Omega_p - \omega_p \right) \Omega r^3 e(r) 
={d  \over d r} \left(
{r^3 c^2\over  \Sigma}{d [\Sigma e(r) ]\over dr}\right)
- {d\left( r^2\Phi'\right) \over  d r}
,\label{ENM}\ee
where $,c,$ is the local sound speed. 

In addition we have an equation for each protoplanet
of the form

\be 2\left(\Omega_p - \omega_p \right) \Omega(r_p) r_p^3 e_p
= - \left[{d\left( r^2\Phi'\right) \over  d r}\right]_{r=r_p}
,\label{ENMp}\ee
where of course
for a particular protoplanet there is no self-interaction
term in the sum for $\Phi'_{ext}.$                                 
The inclusion of self-gravity in the eigenvalue problem
is essential if modes with prograde precession frequency
are to be obtained.
For typical protoplanetary disc models, self-gravity
can be strong enough to induce prograde precession
for the long wavelength $m=1$ modes considered here.

Normal modes were calculated by discretizing (\ref{ENM}) and (\ref{ENMp})
and formulating a matrix eigenvalue problem on an unequally 
spaced grid with $200$ grid points with intervals
increasing in geometric progression  (see Terquem \& Papaloizou 2000
for  consideration of a  related problem).  
 
\subsection{ Disc models}
We present here results obtained for a disc model 
with equation of state given by a polytrope of index $n=1.5.$
The sound speed is given by
\be c^2 ={GM_* H^2\over R_{in}^3}  \left(1-\left(R_{in}/r \right)^{10}\right)
\left(1-\left( r/R_{out}\right)^{10} \right) R_{in}/r.\ee
Here the first two factors provide sharp
edges at the disc inner and outer boundaries  and
$H/r= 0.05$ is the disc aspect ratio away from these boundaries.

The surface density is given by
\be \Sigma = \Sigma_0 (c^2)^n .\ee

The boundary regions are chosen to be 
of order the disc scale height in width
and away from these $\Sigma \propto r^{-3/2}.$
In the above $R_{in}$ is the inner boundary radius which is taken to
be the unit of length. The unit of mass is the central mass $M_*$
and the unit of time is $R_{in}^{3/2}/\sqrt{GM_*}.$
The arbitrary scaling factor $\Sigma_0$ was chosen
such that in model $A$ the  total disc  mass was
$4.0\times 10^{-2}M_*.$ In model $B$ the total disc mass
was $4.0 \times 10^{-3}.$ 
 $R_{out}$ is the outer boundary radius, here,
in order to study large scale modes, taken as $100R_{in}.$

The importance of self gravity is measured by the 
Toomre parameter $Q=\Omega c/(\pi G \Sigma).$
For model A this has a minimum value of $5.2$ while for
the lower mass disc model of model B this minimum value
is $52.$  Another quantity of interest is the local precession frequency
of a test particle orbit $\omega_{pg}$ in the axisymmetric
component of the total   gravitational
potential ( see equation (\ref{wpg}) below).
\begin{figure}
\epsfig{file=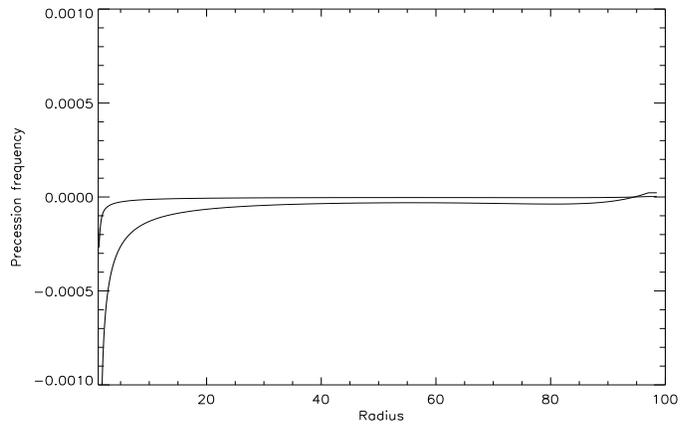, height=6cm,  width=9cm, angle=360}
\vspace{2mm}
\caption{This figure shows the local test
particle orbital  precession frequency in dimensionless units
as a function of  radius, $r$  in units of $R_{in}$
in the range $1.1 <r< 100$
for the two disc  models A (lower curve) and B
with no interior orbiting protoplanets. Apart from  small regions 
near to the boundaries, the  frequency is negative
corresponding to  retrograde precession.}
\label{fig01}
\end{figure}                        
This  is plotted for the two disc models in figure \ref{fig01}.
It is small and negative
 in dimensionless units over much of the discs,
corresponding to retrograde precession,
scaling with the disc mass as it is determined by the disc self-gravity.
As we shall see this behaviour provides scope for secular
resonance associated with low frequency modes.                          
                                             
\section{Normal modes} \label{s2}

We have calculated the lowest order global normal modes 
for two different configurations involving protoplanets
orbiting interior to the disc in an inner cavity.
The first, approximating the Upsilon Andromedae
system  involves two protoplanets with mass ratios $0.00383$
and $0.00196$ orbiting at $0.6R_{in}$ and  $0.194R_{in}$
respectively. However, the disc normal modes do not change much
in character if the protoplanet orbital radii are scaled
to somewhat smaller fractions of $R_{in}.$ But the ratio of protoplanet
to disc eccentricity in the joint modes changes more significantly.
The second configuration we consider is
a single protoplanet with mass ratio $0.002$ orbiting at $0.6R_{in}.$
Finally we consider a  disc with no interior orbiting protoplanets.
In all of these cases
we consider both model $A$ and model $B$ discs.

The pattern speeds for the  highest frequency normal modes
and the  protoplanet orbital
eccentricities  occurring jointly with the normal modes
are given in  table \ref{table1}.  
For the results presented here, 
the modes are normalized such that the disc eccentricity at the inner
boundary is 0.1. 

\begin{figure}
\epsfig{file=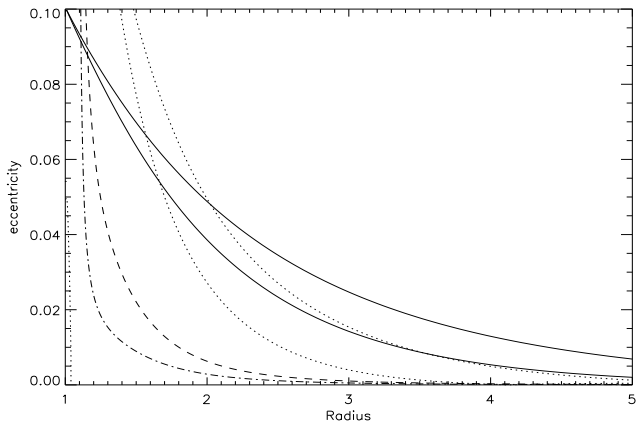, height=6cm, width=9cm, angle=360}
\vspace{2mm}
\caption{This figure shows the first two modes
and the associated  equilibrium eccentricities
as a function of radius $r$ in units of  $R_{in}$
for the two protoplanet  case with disc model A.
The two modes (full curves)
are normalized  so that $e=0.1$ at $r=1.$ The lower curve
corresponds to the mode of highest frequency given in  table \ref{table1}.
The  dotted curves give the associated equilibrium eccentricities
 calculated for a low mass
protoplanet orbiting within the disc,
the lowermost corresponding to the highest frequency mode.
 The dashed curve gives the equilibrium eccentricity for a low mass
protoplanet orbiting within the disc calculated for the
highest frequency mode but neglecting the non axisymmetric component
of the disc potential. The dot dashed curve
gives the corresponding plot for the other mode.
The latter curves indicate the presence 
of secular resonances at $r \sim 1.1$
 associated with both modes.}
\label{fig1}
\end{figure}

\begin{figure}
\epsfig{file=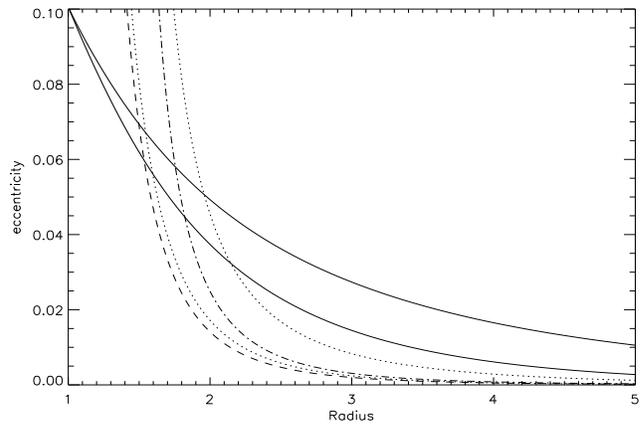, height=6cm, width=9cm, angle=360}

\caption{
 As in figure \ref{fig1}  but for disc model  B.
 In this case there is a secular resonance asociated with the
 highest frequency mode at $r \sim 1.1$. For the other mode
 this occurs at $r \sim 1.4.$}
\label{fig2}
\end{figure}

Many of the properties of these modes can be
understood   with reference to the local
dispersion  relation for density waves in the low frequency limit
(Lin \& Shu 1969) in the form
\be 2\Omega(\omega_p -\Omega_p) = - 2\pi G \Sigma |k| +c^2k^2 ,\ee
where $k$ is the radial wavenumber.
The above indicates that when self-gravity, governed
by the first term on the right hand side,  is unimportant
as occurs for either low mass discs or large $|k|,$ $\Omega_p$ is negative.
On the other hand positive $\Omega_p$ can occur for low $|k|$
or massive discs. Thus in that case the lowest order modes (in terms of numbers of nodes)
should be prograde with more of these existing for more
massive discs, a trend we find in our results.
There can also be very low frequency (in magnitude) modes for which
the self-gravity and pressure terms approximately cancel.
This  occurs when $|k| = 2\pi G \Sigma/c^2 \sim 2/(QH).$
For $Q \sim r/H$ as in the models here, this is comparable
to the radius making a very global mode.

As we calculate normal modes jointly involving protoplanets and disc,
the protoplanets have associated eccentricities.
In general the ${\cal N}$  highest frequency modes predominantly
involve the protoplanets while the others predominantly
involve the disc,  ${\cal N}$ being the number of protoplanets.
The two  highest  frequency modes for the two protoplanet
case are plotted in figure \ref{fig1} for disc model $A$
and in figure \ref{fig2} for disc model $B.$                     
For the second highest frequency mode  calculated for
disc model $A$, the protoplanet
eccentricity ratio is $1.38,$ while for model $B$ it is $1.88.$ 
The latter result is similar to that given by
Chiang, Tabachnik, \& Tremaine (2001) who considered the
current Upsilon Andromedae system with no disc and
concluded it was predominantly in this mode.

\begin{figure}
\epsfig{file=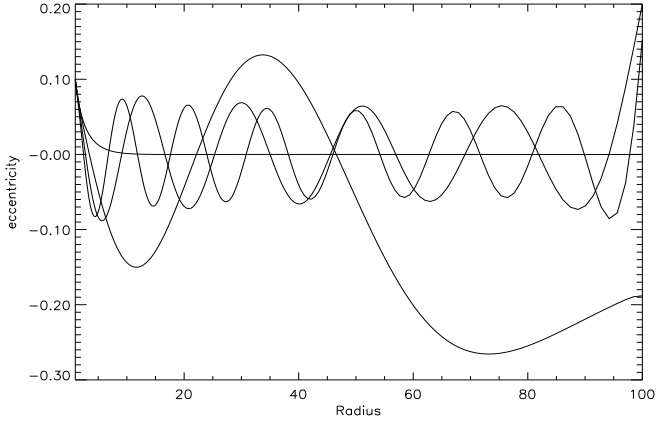, height=6cm, width=9cm, angle=360}

\caption{
This figure shows the  four highest frequency
modes
as a function of  radius $r$ in units of  $R_{in}$
for the one  planet model with disc A. The eccentricities
are all normalized   so that $ e = 0.1$ at $r=1.$
The modes can be identified by noting that the number of nodes
increases with decreasing eigenfrequency.
\label{fig3}}
\end{figure}
	 
\begin{figure}
\epsfig{file=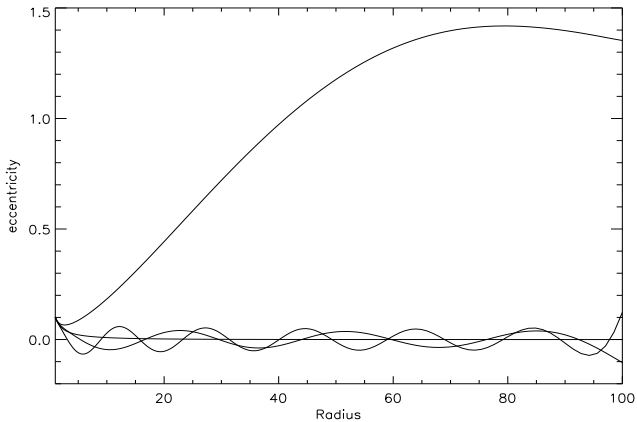, height=6cm, width=9cm, angle=360}

\caption{
 As in figure \ref{fig3} but for the one planet
model with disc model $B.$
\label{fig3b}}
\end{figure}

The results found here suggest the disc and protoplanet
orbits were antialigned. With the outermost protoplanet
at $0.6R_{in}$ the disc inner boundary
and outer protoplanet eccentricities are comparable
for model $A$ while for model $B$ the relative disc eccentricity
is $56$~percent smaller.
However, when the outer protoplanet orbits at
 $0.5R_{in}$ the disc inner boundary
  eccentricity is only $1/3$ that
of the protoplanet for model $A$ and $19$~percent for model $B.$
Apart from the eccentricity scaling relative to the
interior protoplanets, the spatial form
of the eigenfunction in the disc remains almost identical.

Apart fom the two modes with highest
frequency,  other normal modes are essentially pure disc modes with only
small associated protoplanet eccentricities.  

\begin{figure}
\epsfig{file=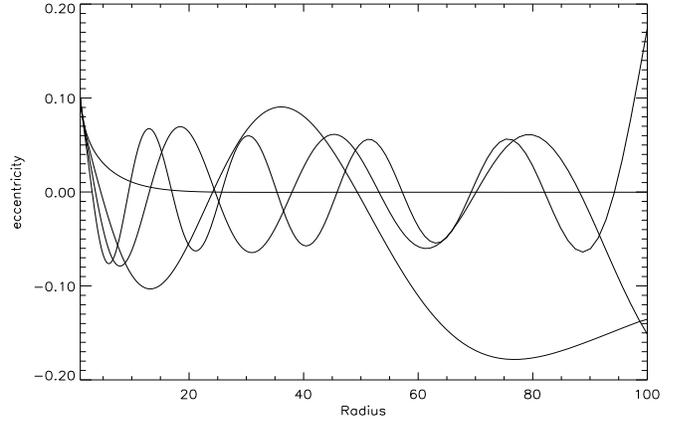, height=6cm, width=9cm, angle=360}

\caption{\label{fig4}  As in figure \ref{fig3} but
for the   disc model A with no protoplanets.}
\end{figure}
     
\begin{figure}
\epsfig{file=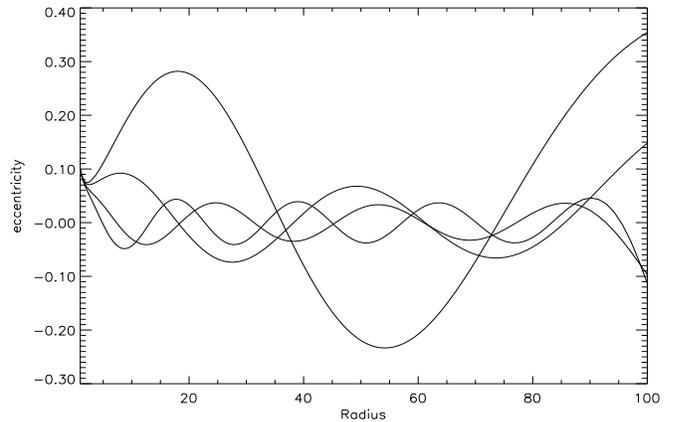, height=6cm, width=9cm, angle=360}

\caption{\label{fig4b}  As in figure \ref{fig4} but
with disc model B}
\end{figure}

The four highest frequency modes for the one protoplanet
case with disc model $A$ are plotted in figure \ref{fig3}
and in figure \ref{fig3b} for model $B.$
Pure disc modes
with no protoplanet are plotted in figure \ref{fig4}  for model $A$
and figure \ref{fig4b}  for model $B$. The modes develop increasing
numbers of nodes as their frequency decreases but the mode
with lowest frequency in absolute  magnitude can be very global
with only a few nodes out to $r=R_{out}.$
We comment that the modes are rather non compressive
requiring eccentricities  of order unity to provide Lagrangian
changes in surface density of order unity. As the motion in the modes
in linear theory
is assumed epicyclic, this is suggestive that the analysis should be valid
as long as the epicyclic approximation is.

\begin{figure}
\epsfig{file= 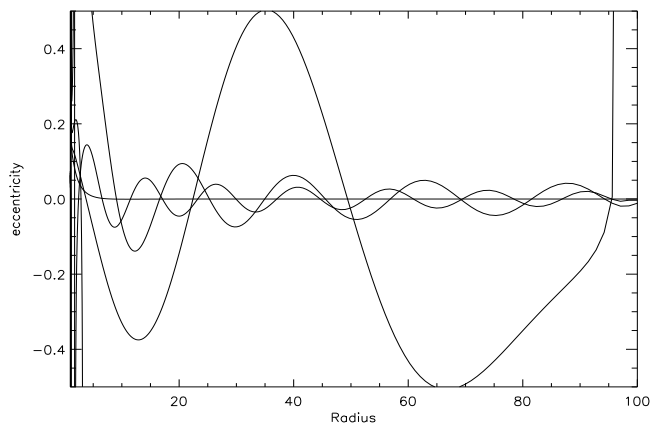, height=6cm, width=9cm, angle=360}

\caption{\label{fig5} This figure shows the  form of the
equilibrium  eccentricity
as a function of  radius $r$ in units of  $R_{in}$
for the modes shown in  figure \ref{fig3},
the latter being  normalized so that $e=0.1$ for $r=1.$
Curves are associated with normal modes according to
increasing number of nodes. Thus the curve with the most
nodes is associated with the normal mode
with the most nodes.  Note the secular resonances for $r <5.$ }
\end{figure}
      
\begin{figure}
\epsfig{file=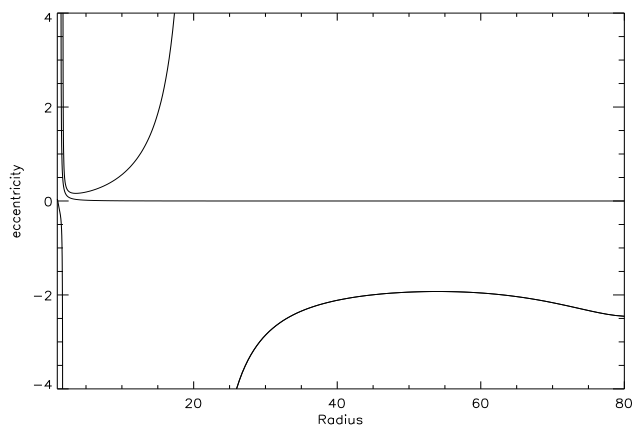, height=6cm, width=9cm, angle=360}

\caption{\label{fig5b}
As in figure \ref{fig5} but for the
one planet model with disc model $B.$
Equilibrium curves are given for the two highest frequency
modes only. That associated with the highest frequency
decays to small values at about $r=5.$
The other curve shows a strong secular resonance
at $r \sim 20.$}
\end{figure}

Having calculated these modes that may be largely driven by protoplanets
in eccentric orbits or  possibly self-excited  or long lived structures
we now go on to consider the migration of low mass protoplanets
embedded in such eccentric discs.

\begin{table}
 \begin{center}
 \begin{tabular}{|l|l|l|l|l|l|l|l} \hline \hline
Disc &$m_{p1}$&$m_{p2}$&$e_{p1}$&$e_{p2}$&$\Omega_p$\\
 \hline
A&$.00383$&$.00196$&$-.36$&$1.82$&$1.61\times 10^{-3}$\\

A&$.00383$&$.00196$&$-.09$&$-.065$&$6.99\times 10^{-4}$
\\
A&$.00383$&$.00196$&$.0068$&$.0029$&$3.01\times 10^{-5}$   
\\
A&$.00383$&$.00196$&$.0035$&$.0011$&$-5.10\times 10^{-4}$    
\\
B&$.00383$&$.00196$&$-.53$&$3.59$&$1.53\times 10^{-3}$                 
 \\
B&$.00383$&$.00196$&$-.18$&$-.094$&$3.58\times 10^{-4}$
\\
B&$.00383$&$.00196$&$.0013$&$.00054$&$-6.53\times 10^{-6}$         
\\
B&$.00383$&$.00196$&$.0006$&$.0019$&$-4.40\times 10^{-4}$           
\\

A&$.002$& - &$-0.091$& -  &$4.00\times 10^{-4}$
\\
A&$.002$& - &$0.0136$& -  &$4.02\times 10^{-6}$
\\
A&$.002$& - &$0.0060$& -  &$-4.06\times 10^{-4}$   
\\
A&$.002$& - &$0.0034$& -  &$-9.17\times 10^{-4}$                 
\\     
B&$.002$& - &$-0.255$& -  &$3.80\times 10^{-5}$
\\
B&$.002$& -  &$0.0122$& -  &$6.62\times 10^{-6}$
\\
B&$.002$& - &$0.0013$&  - &$-3.55\times 10^{-4}$
\\
B&$.002$& - &$0.0005$&  - &$-9.70\times 10^{-4}$
\\   

A&   -   & - &  -&  - &$1.32\times 10^{-4}$
\\
A&   -    & - & - & -  &$3.26\times 10^{-6}$
\\
A&   -    & - & - & -  &$-1.58\times 10^{-4}$
\\
A&   -    & - & - & -  &$-4.12\times 10^{-4}$
\\              

B&   -   &  -&  -&  - &$-6.15\times 10^{-5}$
\\
B&   -    & - & - & -  &$-1.60\times 10^{-4}$
\\
B&   -    & - & - & -  &$-3.74\times 10^{-4}$
\\
B&  -    &  -& - & -  &$-5.77\times 10^{-4}$ 
\\
 \hline
 \end{tabular}
\end{center}
 \caption{This table  gives the  protoplanet orbital
eccentricities where appropriate  together
with the pattern speeds for the  normal modes
 calculated.  The disc model used is indicated in the first column,
the protoplanet mass  ratios to the central star
in the second and third columns, their
orbital eccentricities in the fourth and fifth column and 
the mode frequency or  pattern speed in the sixth column.
The modes are normalized such that the disc eccentricity at the inner
boundary is 0.1. A negative protoplanet eccentricity then indicates
the orbit is antialigned with the disc.}
 \label{table1}
 \end{table}           

\begin{figure}
\epsfig{file=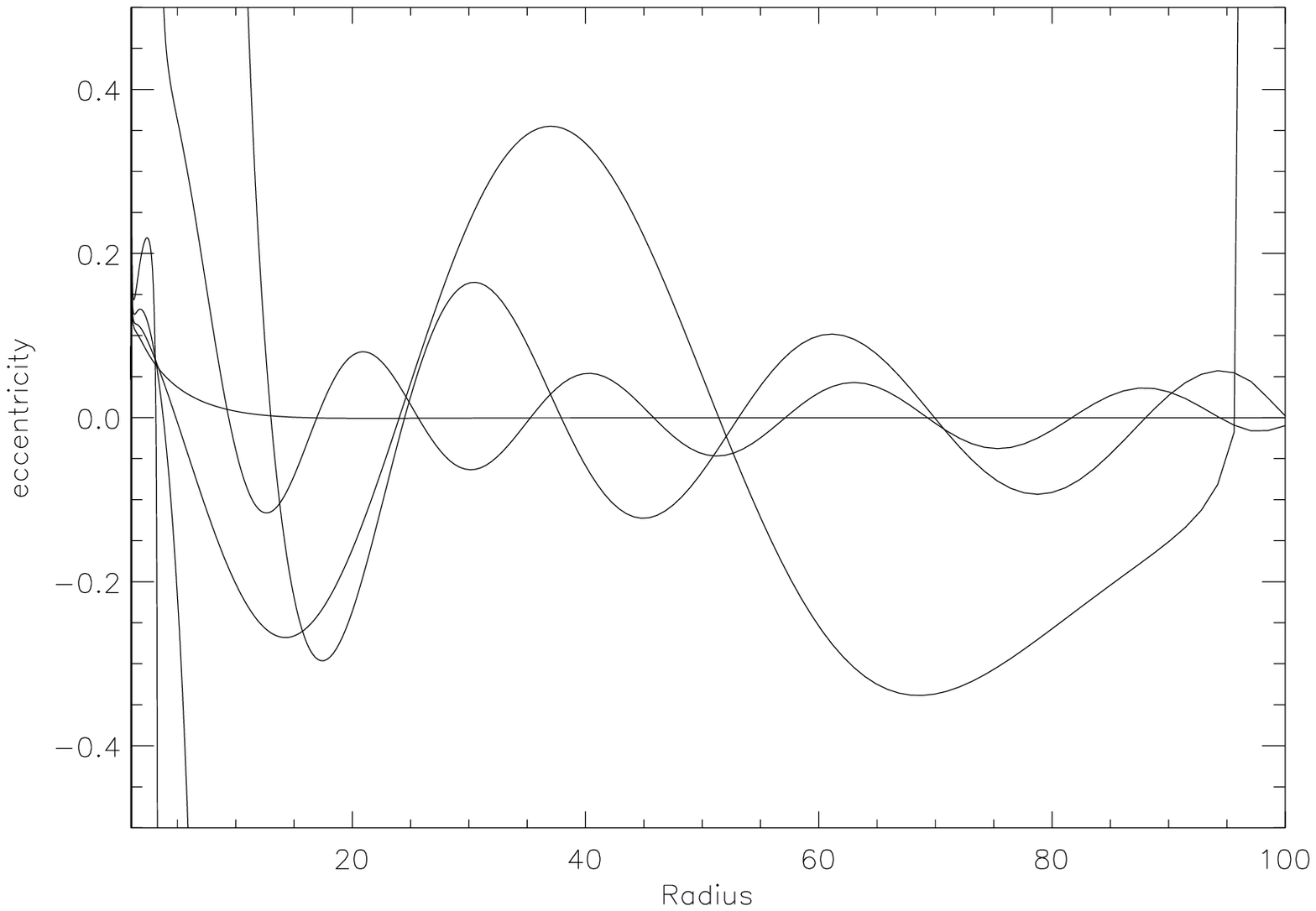, height=6cm, width=9cm, angle=360}

\caption{\label{fig6}
 As in figure \ref{fig5} but for the  disc model $A$
with no protoplanets.}
 \end{figure}

\begin{figure}
\epsfig{file=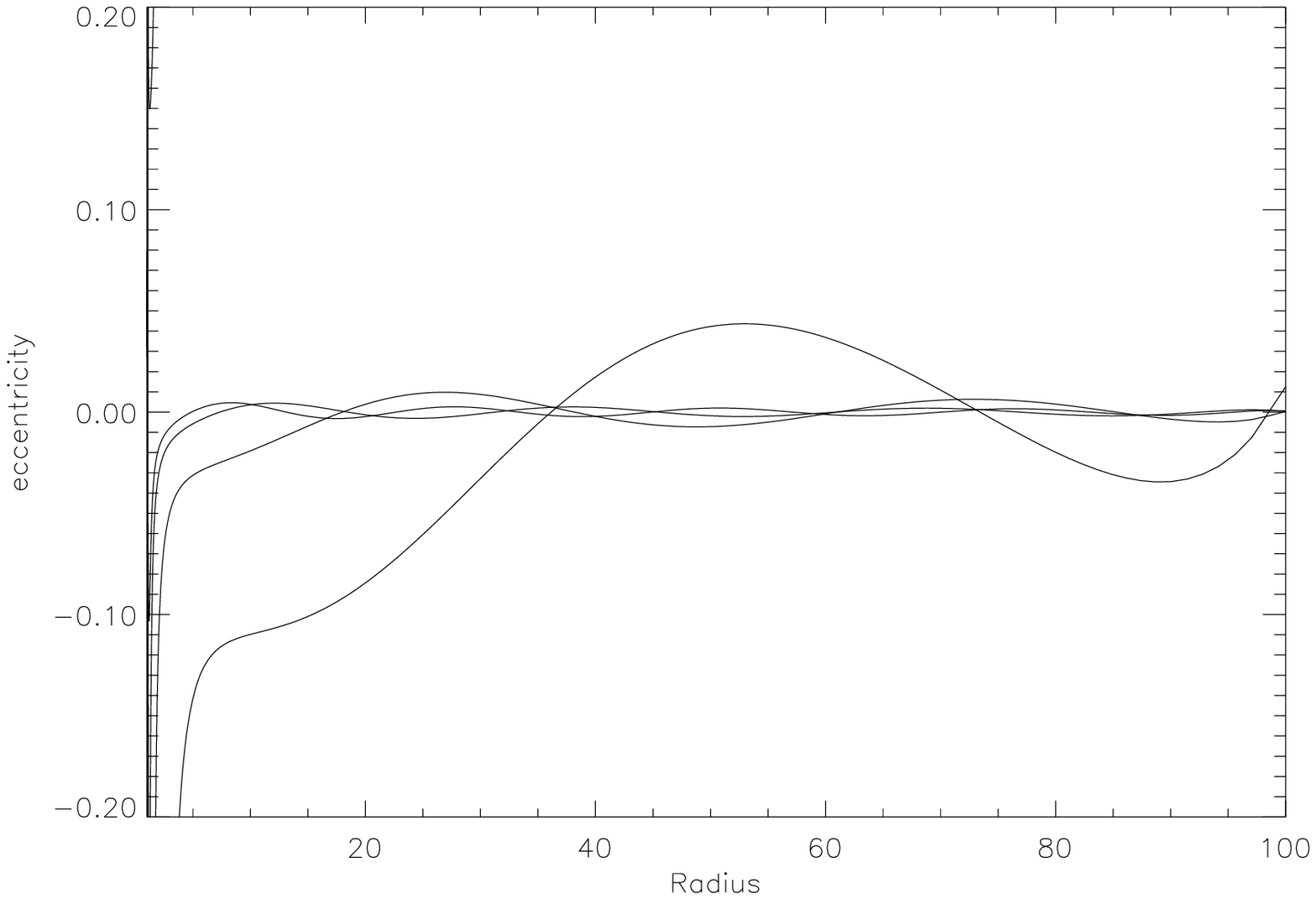, height=6cm, width=9cm, angle=360}

\caption{\label{fig6b}
 As in figure \ref{fig6} but for the  disc model $B$
with no protoplanets.  In this case the longest wavelength equilibrium
eccentricity indicates antialignment with the disc.}
 \end{figure}

\section{ The motion of a low mass
protoplanet in an eccentric disc} \label{s3}
The  evolution of the coplanar  orbit of an interior protoplanet
in the earth mass range embedded in the disc
due to the gravitational interaction
with the slowly precessing disc can be found using secular perturbation theory.
We regard the embedded protoplanet as  a test particle
evolving in a prescribed
gravitational potential and neglect the influence of the protoplanet on the
disc mode. This should be reasonable  when as here, the angular momentum
of the protoplanet is significantly less than that of the material supporting
the normal mode ( eg. Papaloizou et al 2001).

The orbit evolves under the perturbing potential per unit mass,

\be {\cal{ V}} \equiv \Phi_D + \Phi_{ext}  =  \Phi_0(r)
+\Phi_1(r)\cos(\varphi -\Omega_p t).\ee
Here $\Phi_0(r)$ is the axisymmetric component of the potential
but excluding that due to the central mass and $\Phi_1(r)$ is the radial
amplitude of the $m=1$ component arising from the normal mode.

\noindent On performing a time average over the protoplanet orbit
 one obtains the Hamiltonian system:

\be {dh\over dt} =  -{\partial {\cal{H}}\over \partial \alpha}, \ \ \ \  {d \alpha \over dt} ={\partial {\cal{H}} \over \partial h }. \ee

Here $\alpha= \varpi  -\Omega_p t,$ with $\varpi $ being the longtitude of
\noindent periapse  and 
 ${\cal{ H}} ={\cal{V}} -\Omega_p h .$
The specific angular momentum is 
$h= \sqrt{GM_* a_p(1-e_p^2)},$  
$a_p$ is the constant semi-major axis,
and $e_p$ is the eccentricity.
To leading powers of  $e_p$ we have 
\be {\cal{H}} =  A(a_p)+ B(a_p) e_p^2 -\Omega_p h
 +C(a_p)e_p^4 
+ e_pD(a_p)\cos\alpha . \ee

 Hamilton's equations then  give to leading powers of $e_p$
 \be{d\alpha \over dt} = -{2(B+e_p^2(2C-B/2))\over \sqrt{GM_*a_p}}
 -\Omega_p - 
{D(a_p)\cos\alpha\over \sqrt{GM_*a_p} e_p} .\label{ham0} \ee      

\be {d e_p\over dt} = {\sqrt{1-e_p^2}\over \sqrt{GM_* a_p} e_p}
{\partial {\cal{H}}\over \partial \alpha} = 
-{D(a_p) \over \sqrt{GM_* a_p}}\sin\alpha . \label{ham1}\ee  

Here the precession rate of the apsidal line induced by
the axisymmetric potential, to first order in
the perturbing potential, expressed  as an expansion in powers
of  $e_p$ is given by the first term on the right hand
of equation(\ref{ham0}) as 
\be  \omega_{pg} =-{2\over \sqrt{GM_*a_p}}
\left(B+e_p^2(2C-B/2)\right) .\ee

This may also be written in terms of the axisymmetric
component of the  potential directly as
\be 2\omega_{pg}\sqrt{GM_*a_p^5}
=(1+e_p^2){d^2 \Phi_0 \over d u^2} +
{e_p^2\over 8a_p^2}{d^4\Phi_0\over du^4}, \label{wpg}\ee
where $u=1/r,$ and $r$ is taken to be equal to $a_p.$

Equilibrium  or steady state solutions of (\ref{ham0})
and (\ref{ham1})  with $e_p$ and $\alpha$  constant
occur when  $\alpha=0$ or $\alpha=\pi.$
Adopting the convention that $e_p$ is positive suffices
to select one of the latter possibilities as
(\ref{ham0}) gives an expression from which $e_p$
can be determined in the form

\be e_p = -{{D(a_p)\cos\alpha}\over{{2(B+e_p^2(2C-B/2))} 
+\Omega_p \sqrt{GM_*a_p} }}
.\label{ham2} \ee
or equivalently
\be e_p = {{D(a_p)\cos\alpha}\over{(\omega_{pg}
-\Omega_p)\sqrt{GM_*a_p} }}
.\label{ham3} \ee
 Alternatively one may fix $\alpha$ to be $0$ or $\pi$
and allow $e_p$ to change sign as we have done for the normal modes.

An equilibrium solution so determined corresponds to the situation
when the protoplanet orbit precesses at the same rate, $\Omega_p,$  as the mass distribution
that produces the gravitational potential while maintaining a constant
eccentricity. Assuming that $\omega_{pg}$ and
$\Omega_p$ are of comparable magnitude, in general
for modest eccentricity, 
$$e_p \sim D(a_p)/(\Omega_p \sqrt{GM_*a_p}) \equiv e_0$$
and is proportional to the magnitude of the nonaxisymmetric potential.
An exception occurs when 
$${2B
+\Omega_p \sqrt{GM_*a_p} }=0.$$
In this case, the precession frequency
of a free orbit with small eccentricity, $\omega_{pg},$ matches that
of the nonaxisymmetric mass distribution $\Omega_p$ corresponding to a secular resonance.
When this occurs equation (\ref{ham2}) indicates that $e_p \sim (e_0)^{1/3}.$
Accordingly significantly larger equilibrium eccentricities are expected
close to a secular resonance.  We comment that for fixed $\alpha$
$e_p$ changes sign as a secular resonance is passed through corresponding
to an alignment change  through 
a rotation of the axis of the  ellipse through $\pi.$

One may also investigate the effect of orbital circularization
by adding a term $-e_p/|t_e|$ to the right hand side
of (\ref{ham1}), where $|t_e|$ is the circularization time. 
In this case one can still find an equilibrium but with orbital
apsidal line rotated. Restricting consideration to the situation
away from secular resonance, one finds that the equilibrium eccentricity
is reduced by a factor $\sqrt{1+1/((\omega_{pg} -\Omega_p)^2t_e^2)}$
and $|\sin \alpha| = 1/\sqrt{1+(\omega_{pg} -\Omega_p)^2t_e^2}.$
Thus when $|(\omega_{pg} -\Omega_p)t_e|$ is large, 
the effect of the circularization term is to produce
 a small rotation of the apsidal line of the orbit.

As indicated above equilibrium solutions correspond to the situation where
the apsidal precession
of the protoplanet  orbit is locked to that of the underlying nonaxisymmetric disc.

One can find solutions undergoing small librations   in the neighbourhood
of equilibrium
solutions (eg. Brouwer \& Clemence 1961) and when dissipative forces
are added
these may decay making the attainment of equilibrium solutions natural.
As we indicate below, tidal interaction of a protoplanet with the disc may
produce such dissipative forces. Accordingly we shall focus
on equilibrium solutions in what follows below.

\subsection{Determination of equilibrium  eccentricities corresponding
to normal modes}

It is a simple matter to determine equilibrium
eccentricities for protoplanets moving under the gravitational
potential appropriate to a normal mode corresponding
to an eccentric disc, with pattern speed $\Omega_p,$
 of the type calculated above.
We recall the normal mode equation(\ref{ENM})
which can be regarded as determining the equilibrium
disc eccentricity in the form
\be 2\left(\Omega_p - \omega_p \right) \Omega r^3 e(r)
= {d  \over d r} \left(
{r^3 c^2\over  \Sigma}{d [\Sigma e(r) ]\over dr}\right)
- {d\left( r^2\Phi'\right) \over  d r}
 .\label{ENMa}\ee  

If the pressure forces are neglected by dropping
the terms  involving $c^2,$ this is exactly equivalent
to equation (\ref{ham3}) for determining the
protoplanet eccentricity with $e(r)$ corresponding
to $e_p$ and $\omega_p$ replaced by $\omega_{pg}.$
 Thus a local protoplanet equilibrium eccentricity
appropriate to a disc mode or response
can be determined by using the disc  normal mode equation (\ref{ENM})
retaining the term resulting from the disc self-gravity but omitting
pressure contributions.

This  procedure should be applicable provided
the eccentricities are not too large. Equilibrium eccentricities
are  plotted for some of the normal modes we calculated
 in figures \ref{fig1} and \ref{fig2} and also  \ref{fig5} - \ref{fig6b}.
In figures \ref{fig1} and \ref{fig2} we also plot equilibrium eccentricities
but calculated  with the nonaxisymmetric contribution to the gravitational
potential  due to the disc removed  which is equivalent to
assuming that it remains circular. We see that the assumption that
disc remains circular  results in a 
significantly smaller equilibrium eccentricity for disc model A,
particularly for the mode with second highest frequency.
For the lower mass disc model B, differences arising
fom assuming the disc
remains circular are less pronounced. 
We also coment that the occurence of secular resonances
in the inner parts of the disc where $r/R_{in} < 10$
is common.

One expects that  dissipative torques produced by

\noindent protoplanet-disc tidal
interaction  may result in the  approach to such equilibrium
solutions from general initial conditions.  This we now discuss.

\section{Response of an eccentric disc to a low mass protoplanet} \label{s4}
To calculate the tidal
response of an eccentric disc to a perturbing protoplanet it is convenient
to introduce a new orthogonal coordinate system
$(a,\lambda),$ with $a$ being the local semi-major axis and $\lambda$
being the orthogonal angular coordinate.
Then taking  pericentre in the disc to lie along $\varphi =0,$
$a=r(1+e\cos\varphi)/(1-e^2),$
with the eccentricity being a function of $a.$
We assume the eccentricity is small enough
that we may work to first order in it and also that
it has significant changes only on a global length
scale comparable to $r.$  Then we may make the replacement
$e(a) \equiv e(r).$
In this approximation $a$ can be expressed in terms
of cylindrical coordinates in the form $a=r(1+e\cos\varphi)$ and the orthogonal
angular coordinate, $\lambda$ is given  by
\be \lambda= \varphi +\int({e\over r}dr)\sin\varphi,\ee
where we  may leave the integral as indefinite.
It is also convenient to work in a frame in which the disc flow
appears stationary. As the pattern speed as measured in an inertial
frame is very  small compared to the disc rotation speed, we shall
neglect centrifugal and coriolis forces.

In a general orthogonal coordinate system in which the disc
appears to be in a steady state, the equations of motion
for the velocity $( v_{a}, v_{\lambda})$
may  then be written ( see Appendix 2)
$$ {\partial  v_{a} \over\partial t}+
v_{a}|\nabla a|^2{\partial\over\partial a}\left({v_{a}\over |\nabla a|}\right)
+v_{\lambda}|\nabla a||\nabla \lambda|{\partial 
\left({v_{a}\over |\nabla a|} \right)\over\partial \lambda}$$

$$-{v_{\lambda}^2 \over 2} |\nabla a||\nabla \lambda|^2  
{\partial \left(|\nabla \lambda|^{-2}\right) \over\partial a}
-{v_{a}^2 \over 2} |\nabla a|^3 {\partial \left(|\nabla a|^{-2} \right)\over\partial a}$$
\be = -{|\nabla a|\over \Sigma}{\partial \Pi \over\partial a}-
|\nabla a|{\partial \Phi \over\partial a},
\label{12d1} \end{equation}

 $$ {\partial  v_{\lambda} \over\partial t}+
 v_{\lambda}|\nabla \lambda|^2{\partial \left(
{v_{\lambda}\over |\nabla \lambda|} \right)\over\partial \lambda}
+v_{a}|\nabla a||\nabla \lambda|{\partial \left(
{v_{\lambda}\over |\nabla \lambda|} \right)\over\partial a}$$
$$-{v_{a}^2 \over 2} |\nabla a|^2|\nabla \lambda|  {\partial \left(
|\nabla a|^{-2} \right)\over\partial \lambda}
-{v_{\lambda}^2 \over 2} |\nabla \lambda|^3 
{\partial \left(|\nabla \lambda|^{-2}\right) \over\partial \lambda}$$
\be \ \ \ \ \ = -{|\nabla \lambda|\over \Sigma}{\partial \Pi \over\partial \lambda}-
|\nabla \lambda|{\partial \Phi  \over\partial \lambda},
\label{12d2} \end{equation}

We perform a response calculation by linearizing about a steady state in which
the disc gas  is taken to be  in elliptical keplerian orbits with 
$v_
a =0,$
and to first order in $e,$  $v_{\lambda}=(GM_*/a)^{1/2}(1+e\cos\lambda).$
In a thin disc forces due pressure and self-gravity
provide only a small correction of order $(H/r)^2.$ 
We consider the situation when the disc eccentricity, $e(r),$ protoplanet eccentricity, $e_p,$
and $H/r$
can be considered small and  comparable in magnitude.

When considering the response to an embedded protoplanet, we are interested in responses with a scale
$H$ and azimuthal mode number $m \sim r/H.$  Here for the time being $\lambda$ is regarded as the azimuthal angle.
We perform a response calculation by linearizing
equations (\ref{12d1}) and (\ref{12d2}) about the  steady state 
described above.
We may  expand the linearized equations in powers of the disc eccentricity.

Here we shall work only to lowest order and neglect terms of order $e$ times smaller than the dominant ones
which,  linearizing (\ref{12d1}) and (\ref{12d2}) directly and 
denoting perturbation quantities by a prime, are of order $\Omega v_a'.$
In this  scheme we may replace $|\nabla a|$ and $|\nabla \lambda |$ by unity except where the latter
occurs in the combination 
$ |\nabla \lambda |v_{\lambda}{\partial \over \partial \lambda},$  
with $|v_{\lambda}|$ here denoting the unperturbed velocity in the disc.
This is because  the  contribution  of terms of first order in the eccentricity to this operator
leads to
quantities of order $m e |v_{\lambda}| v_a'/r$ in the linearized equations.
For $m \sim r/H, \sim 1/e $ these are comparable to the dominant terms. All other eccentricity
contributions are smaller by a factor at most $\sim e.$

We have  to first order in eccentricity
\be  v_{\lambda}|\nabla \lambda | =
(GM_*/a^3)^{1/2}\left( 1+ 2 e \cos\varphi + \int({e\over r}dr)\cos\varphi \right) .\ee

Thus if we define a new angle $M$ through $M= \lambda - (2e+\int({e\over r}dr)\sin\lambda,$
the operator $|\nabla \lambda |v_{\lambda}{\partial \over \partial \lambda},$

\noindent becomes $ =(GM_*/a^3)^{1/2}{\partial \over \partial M}.$
The angle $M$ is just the  mean anomaly as can be seen from the fact that to first
order in $e,$ 
\be r=a(1-e\cos M), \varphi= M+2e\sin M, \label{PPO}\ee  For the motion
of  a particular disc particle we have
$M=\Omega(t-t_0),$ where $t_0$ denotes the time of periastron passage.

To lowest order in $e,$ the coordinates $(a,M)$ behave just like the cylindrical coordinates
$(r,\varphi)$ in the sense that one may make the replacements
$r \rightarrow a, \varphi \rightarrow \lambda $ in the standard  linearized  equations
expressed in cylindrical coordinates.
 However, recall that in the forcing potential $\Phi_p',$ we must  make the replacement (\ref{PPO})
which can be thought of evaluation on  an eccentric disc orbit.

\subsection{Protoplanet forcing potential}
Neglecting the indirect term, which does not contribute
to the  Fourier components with large $m \sim r/H$
of interest, the protoplanet forcing potential is
\be \Phi_p' = -{Gm_p\over  \sqrt{|r^2+R^2-2rR\cos(\varphi -\nu) +b^2|}}
\label{fopo}.\ee
Here, the coordinates of the protoplanet are $(R,\nu)$ and $b$ is  the softening
parameter which is introduced to take account of the  finite vertical thickness
of the disc. In general $b \sim H$ ( eg. Artymowicz 1993).

For an eccentric protoplanet orbit with semi-major axis $a_p$ and eccentricity $e_p,$
 we have
\be R=a_p(1-e_p\cos (\omega t )), 
\nu= \omega t +\varpi +2e_p\sin(\omega t).
\label{snm}\ee
where the mean motion is $\omega ,$ we have taken pericentre passage to
be at $t=0,$  and $\varpi$ is the longtitude of pericentre which is
not necessarily zero as 
we allow for 
the apsidal line of the  orbit not to
be lined  up with that of the disc.

Using (\ref{PPO}) and (\ref{snm}), we perform the Fourier
decomposition 

 $$\Phi_p' =\sum_{m,n,n_1} C_{m,n,n_1}
\exp i\left((n-m)\omega t + m(M-\varpi)+n_1M\right).\label{FDEC}$$

Here, the sum is over positive values of $m$
and both positive and negative values of $n$ and $n_1.$
The convention is that the real part is to be taken to give
the physical potential. The coefficients $C_{m,n,n_1}$
are real and given by
$$ 
{4 \pi^3  C_{m,n,n_1} \over Gm_p \omega} =  \hspace{5cm} $$
\be  -\int
{\exp-i\left((n-m)\omega t - m\varpi+(n_1+m)M\right)
 \over \sqrt{|r^2+R^2-2rR\cos(\varphi -\nu) +b^2|}}
dM d\varpi dt.\ee
The integral is over a $2\pi$ cycle in each of the
angles $\varpi, M, \omega t.$
Note that using (\ref{PPO}) and (\ref{snm})
to express the disc and protoplanet coordinates
makes $C_{m,n,n_1}$ a function of $a,e,a_p,$ and $e_p.$

An important aspect of the linearized problem  
expressed in terms of the coordinates $a,M$ is that
the separable coordinates are $t$ and $M.$
Accordingly a  separable response  occurs to a combination of terms
in the   Fourier decomposition with fixed
$(n_1+m) =k_1$ and $(n-m)=k_2$.

Thus for  such a particular forcing term with fixed $(k_1,k_2)$
\be \Phi_p' = f_{k_1,k_2}
\exp i\left(k_2\omega t +k_1 M\right),\label{FSEP}\ee
where
\be  f_{k_1,k_2} = \sum_{m,n,n_1} \delta_{n_1+m,k_1}  \delta_{n-m,k_2}
C_{m,n,n1} \exp(-im\varpi)\label{COEFF}. \ee
Here $\delta$ denotes the Kronnecker delta.

A novel feature is that, unlike in the case of a forced axisymmetric disc,
a separated forcing
potential component in general  depends on $\varpi.$
This is because the angle between the apsidal lines  of 
the  disc and protoplanet orbits   
can have physical significance. But note that when both protoplanet
and disc eccentricity are zero only one term can survive 
in (\ref{COEFF}) when $m = k_1 = - k_2$ and then $\varpi$ appears
only as a redundant complex phase.

For the general forcing term (\ref{FSEP}) the pattern speed
$\Omega_p = -k_2\omega /k_1.$ We shall consider the situation when
the disc surface density $\Sigma \propto r^{-3/2}$ as is the
case in our models away from the boundaries. Then corotation
resonances may be ignored and  the main interaction
is through wave  excitation at the Lindblad resonances
(eg. Goldreich \& Tremaine 1978). These occur when $k_1(\Omega -\Omega_p)
= \pm \Omega \sqrt{1+\xi^2}$ with $\xi = k_1c/(r\Omega)$ (eg. Artymowicz 1993).  
For $k_1 >0,$ the positive sign applies  to the outer Lindblad resonance (OLR)
and the negative sign to the inner Lindblad resonance (ILR).
In both cases a wave is excited that propagates away from the protoplanet.
The waves are associated with an outward energy and angular momentum flux.
In the case of the ILR,  the background rotates faster than the wave
so that as it dissipates, energy and angular momentum
are  transferred to the protoplanet orbit.
In the case of the OLR, the background rotates more slowly
so that energy and angular momentum are removed from the protoplanet orbit.

Because the linearized response
problem is formally identical to the one obtained for
an axisymmetric disc $(a \rightarrow r, M \rightarrow \varphi)$ ,
consistent with he other approximations
we have made, we can  evaluate the outward energy flow rate associated 
with outward propagating waves
using the approximate
expressions developed by Artymowicz (1993) and  Ward (1997)  which
depend only on the disc state variables evaluated at the 
respective resonances. We comment that there are uncertainties
associated with the use of these expressions in evaluating orbital
evolution rates especially when they are derived  by summing
torque contributions with varying signs.  However, cancellation
effects  do not  appear to be significant 
(Artymowicz 1993,  Ward 1997) and as they are supported by general
considerations, we believe that the  main features of the results
derived to be correct. Following this procedure,
the  outward energy flow rate associated
with outward propagating waves is given by

\begin{equation}
{d E_d\over dt} ={\pi^2
\Sigma  r^2
\over 3\Omega \sqrt{1+\xi^2}(1+4\xi^2) }\left|
{\cal{E}}_{k_1,k_2} \right|^2
,\label{EFF2}
\end{equation}
where
$$ {\cal{E}}_{k_1,k_2} = {d f_{k_1,k_2}\over dr}+
{2m^2(\Omega-\Omega_p)f_{k_1,k_2} \over \Omega r}.$$

However, unlike in an axisymmetric disc, the outward  angular momentum flow 
 in the disc associated with the protoplanet,\newline
$dJ_d/dt \ne (1/\Omega_p)dE_d/dt .$ 

One can find the outward angular momentum flow rate  produced directly
by the protoplanet
associated with the linear response
using 

\begin{equation}
{d J_d\over dt}
= \int \Sigma' {\partial \Phi_p'\over \partial \varpi} rdrd\varphi,
\end{equation}
the integral being taken over the disc.  Recalculation of the torque formula
then gives

\begin{equation}
{d J_d\over dt}=Re\left({\pi^2
\Sigma  r^2
\over 3\Omega \omega |k_2|
\sqrt{1+\xi^2}(1+4\xi^2) }
{\cal{E}}_{k_1,k_2}{\cal{T}}^*_{k_1,k_2}\right)
,\label{EFF3}
\end{equation}
where
$$ {\cal{T}}_{k_1,k_2} = {d g_{k_1,k_2}\over dr}+
{2m^2(\Omega-\Omega_p)g_{k_1,k_2} \over \Omega r} ,$$
with 
\be g_{k_1,k_2}=\sum_{m,n,n_1} m\delta_{n_1+m,k_1} \delta_{n-m,k_2}
C_{m,n,n1} \exp(-im\varpi)\label{COEFF1}. \ee
 
\noindent The rate of change of protoplanet
eccentricity, $e_p$, induced by each term can be found
by using the orbit equation
 
\begin{equation}
{e_p \omega J\over (1-e_p^2)^{3/2}}{ de_p\over dt}
= { dE\over dt} -\omega { dJ\over dt}
(1-e_p^2)^{-1/2} 
, \label{ECCEV}
\end{equation}

\noindent where $E,$ and $J$ are the  orbital energy and
angular momentum of the protoplanet 
and $ dJ/dt =  \pm dJ_d/dt, dE/dt = \pm dE_d/dt, $
with the positive sign applying for an ILR and the negative
sign for  an OLR.
 
\noindent The total rate of change is found by summing
contributions from each resonance 
 occurring for all values of  $(k_1,k_2).$
 
\noindent The eccentricity evolution time-scale,
which might be associated either with excitation, when it is positive,
 or
damping, when it is negative,  is then
 
\begin{equation}
t_e = \frac{e_p}{{\rm d}e_p/{\rm d}t}.
\end{equation}
 We define the  time for the angular momentum to decay
by a factor of $e$ as
\begin{equation}
t_{J} = - \frac{J}{{\rm d}J/{\rm d}t},
\end{equation}       
such that a positive value means inward migration.
\section{Tidal torques in an eccentric disc}
We have used  the above formalism to estimate
eccentricity excitation/damping rates for protoplanets in eccentric
discs  which can be described  using the normal modes calculated above.
Before giving details we give a brief summary of our results.
Although we consider eccentricities which may substantially
exceed $H/r,$ we shall still suppose them sufficiently small $\le \sim 0.2$
that we may consider there to be an equilibrium eccentricity
as a function of disc radius. Then a torque calculation
is characterized by a disc eccentricity , $e,$ and an equilibrium eccentricity
which we may also consider to be the protoplanet eccentricity
$e_p.$ We shall also for the most part 
restrict consideration to the case
when the protoplanet and disc orbits are aligned in equilibrium
$(\varpi =0.)$  In general when $e_p \ll e$  we find 
excitation of the protoplanet orbit eccentricity, $t_e >0,$
while for $e_p \gg e,$ the eccentricity damps as expected, $t_e <0.$
The transition between these regimes 
occurs when $e$ and $e_p$ are approximately equal.
For $e_p $ significantly larger than $e$,
provided $|(\Omega_p - \omega_{pg}) t_e|$ 
 is significantly  greater than unity,
there is an equilibrium solution with apsidal line slightly rotated
from zero. The orbit may then suffer significantly reduced
or even    reversed torques  for $e_p$ sufficiently large. 

\begin{figure}
\epsfig{file=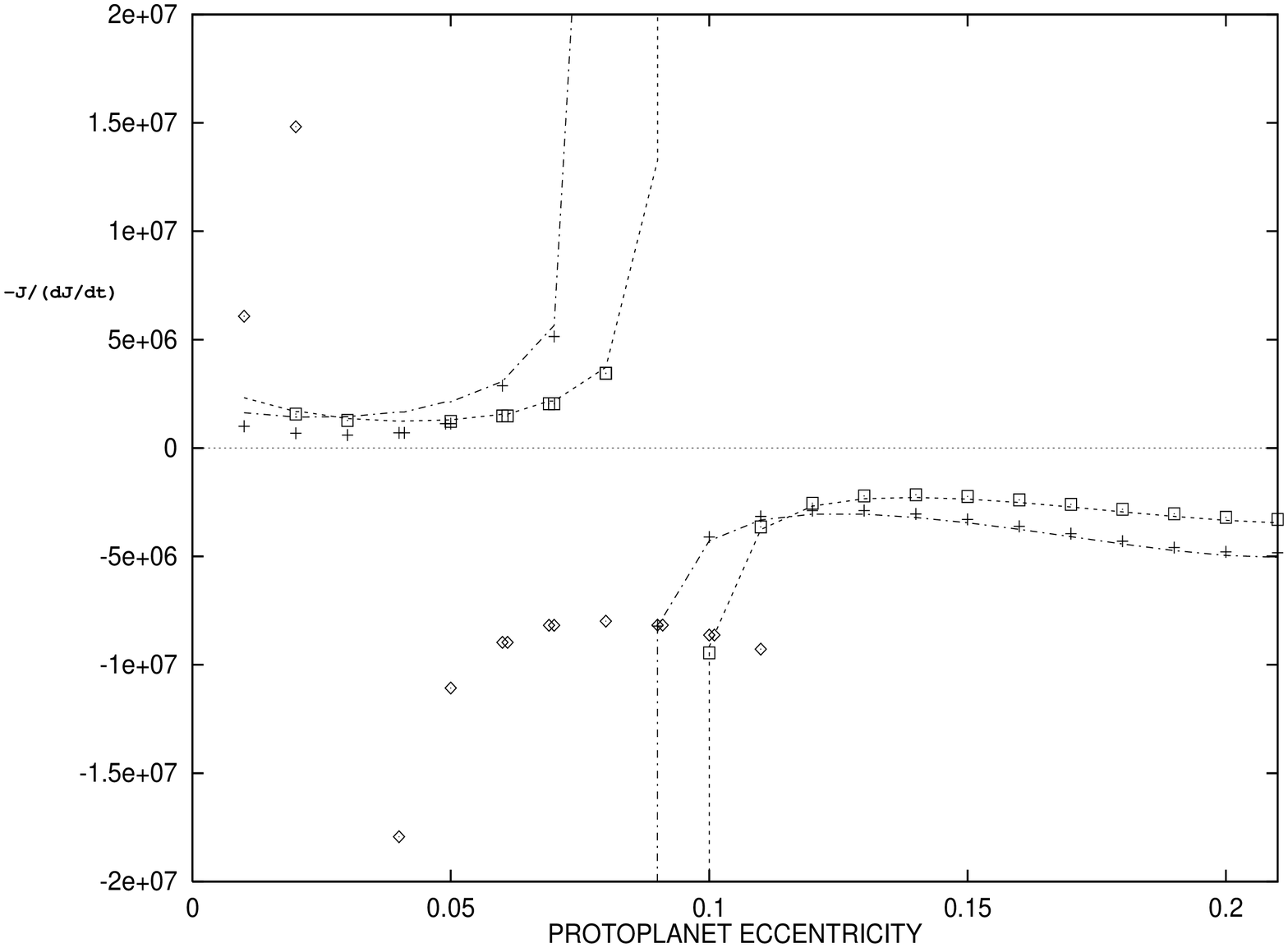, height=8cm,  width=9cm, angle=360}
\vspace{2mm}
\caption{ In this figure $t_{J}$  in yr is plotted 
as a function of protoplanet equilibrium eccentricity
(negative values correspond to  torque reversal ) for $N_t =1.$
The short dashed curves correspond to an assumed disc eccentricity
$e= 0.07$ and the short dashed  long dashed curves to 
$e = 0.05.$ The disc aspect ratio $H/r$  was taken as $0.07.$
The crosses  were obtained for $e=0.05$ and $H/r=0.05,$
while the squares correspond to $e=0.07$ and $H/r =0.05.$
In the above cases  disc and protoplanet orbit
apsidal lines were taken to be aligned. The  diamonds
give $t_{J}$ for assumed antialignment  and $e = H/r = 0.07.$
Note the weakening of the disc interaction at the higher
protoplanet eccentricities due to larger relative velocities
especially in the antialigned case,
to the disc.}
\label{figMIG}
\end{figure}

When $e_p$ is significantly less than a sufficiently
large disc eccentricity $e,$ 
the protoplanet orbital eccentricity can grow  
until the orbit ceases to be aligned with the disc
and precesses through a full $(2\pi)$ cycle
at which point it then damps.
Again inward orbital migration may be reduced or reversed.
Many of these features can be traced to the fact that  when the 
equilibrium eccentricity is such that
$e_p =e$  the situation in many ways replaces the equilibrium eccentricity
solution $e_p=0$ in an axisymmetric disc.  This we show below

\begin{figure}
\epsfig{file=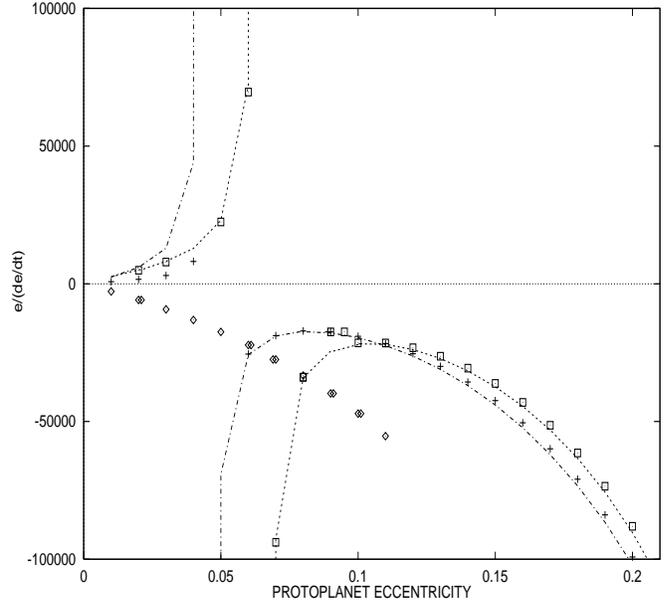, height=9cm,width=8cm, angle=270}
\vspace{2mm}
\caption{ In this figure  $t_{e}$ in yr is plotted 
as a function of protoplanet equilibrium eccentricity
(negative values correspond to circularization) for $N_t =1.$
The short dashed curves correspond to an assumed disc eccentricity
$e= 0.07$ and the short dashed  long dashed curves to 
$e = 0.05.$ The disc aspect ratio $H/r$  was taken as $0.07.$
The crosses  were obtained for $e=0.05$ and $H/r=0.05,$
while the squares correspond to $e=0.07$ and $H/r =0.05.$
In the above cases  disc and protoplanet orbit
apsidal lines were taken to be aligned. The  diamonds
give $t_{e}$ for  an orbit with apsidal line
antialigned  with that of the disc and $e = H/r = 0.07.$
Note the  eccentricity grows for protoplanet eccentricity
less than $e$ in the aligned case while there is
always decay in the antialigned case.}
\label{figCIRC}
\end{figure}

We use the expressions for $r, \varphi, R,$ and $\nu,$
given by equations (\ref{PPO}), (\ref{snm})
in the forcing potential (\ref{fopo}).
In the case when $e = e_p,$ and the protoplanet
orbit is almost aligned with the disc with very small  $|\varpi|,$
the strongest interaction occurs when
$r=R,$ and $\varphi =\nu.$ This corresponds to
$a=a_p,$ and $M= \omega t + \varpi.$ Performing a first order
Taylor expansion about the point of maximum interaction
the forcing potential becomes
\be \Phi_p' = -{Gm_p\over \sqrt{|(a - a_p)^2+4aa_p\sin^2(M -\omega t -\varpi) +b^2|}}
\label{fope}.\ee                   
Here $|a-a_p|$ and $a|M -\omega t|$ are considered
as small compared to $a$ and terms of order $e$ times these
small quantities have been neglected. Equation (\ref{fope})
is to be considered in comparison to the similar
expression appropriate to an axisymmetric disc
\be \Phi_p' = -{Gm_p\over \sqrt{|(r - R)^2+4rR\sin^2(\varphi -\nu) +b^2|}}
\label{fopd}.\ee    
Given that we have already shown that $a,M$ behave just like
cylindrical  coordinates $(r, \varphi)$ the tidal torque calculation
in an eccentric disc with $e_p = e$ should give the same
results as an axisymmetric disc with $e_p =0.$
Note too that only one term should remain in the
sum (\ref{COEFF}) such that $k_1 =m,$ and $k_2 = -m.$
Then there is zero rate of change of eccentricity.
Thus the situation of aligned orbits   such that $e_p = e$
behaves much like the case with $e_p =0$ in an axisymmetric
disc in that it is one of steady eccentricity. The orbit then migrates
at the same rate as a circular orbit in an axisymmetric disc.
This is essentially what we have found on application
of the torque formulae.

\subsection{Numerical results}
We have calculated $t_{J}/N_t$ and $t_e/N_t$ by  summing
the contributions from appropriate resonances. The normalizing
factor 
$$N_t =(r/5 {\rm AU})^{-1/2}(m_p/(3M_{\oplus}))^{-1}(\Sigma/56)^{-1}.$$
Here the distance is measured in units of $5$~AU, the protoplanet mass
in units of $3$ earth masses and the disc surface density
in units of $56 gm cm^{-2}.$ We have also taken the
central mass to be one solar mass.  The softening parameter  was
taken to be $b=H/\sqrt{2}.$ 
Our results are consistent with those of Ward(1997) in the limit
where both $e$ and $e_p$ tend to zero.
 We plot $t_{J}/N_t$  as
a function of equilibrium protoplanet orbit  eccentricity $e_p$  for $e=0.05$
and $e=0.07,$ when disc and protoplanet apsidal
lines are aligned  in figure \ref{figMIG}. Results for  aspect ratios
$0.05$ and $0.07$ are presented. 

We plot  $t_e$  for $N_t=1$  in
figure \ref{figCIRC}. The trends in all cases   are similar
and  are that for small  
 $e_p$ the protoplanet eccentricity grows in the aligned case
while inward migration occurs with $t_{J} \sim 2\times 10^6$~yr.
The eccentricity growth reverses for $e_p \sim e$ indicating
an equilibrium  in accordance with the discussion above.
We comment that for very small $e_p$ and finite $e,$
$de_p/dt \propto e.$
For larger $e_p,$ $t_{J}$ and $t_e$ increase and $t_{J}$
eventually changes  sign for $e_p$ exceeding $\sim 0.1.$
At these eccentricities $|t_e| \sim 2\times 10^4$~yr.
We make the comment that
 the same dimensionless units can be used
for $t_{J}$ and $t_e$ as for the disc models introduced in section 
\ref{s1} and  thus the same scaling to make results
applicable to different radii  may be used.

In the antialigned case the disc protoplanet interaction
is much weaker (see figures \ref{figMIG} and \ref{figCIRC}).
Note too that the interaction with the disc weakens in general
for larger $e_p$ because of the larger relative velocity
of the protoplanet with respect to the disc.
This results in larger values of $|t_e|$ and $|t_{J}|.$
Additional calculations have shown that, as expected, these values
become independent of the orientation of the apsidal
line  when $e_p \gg e.$

Thus the indications are that for modest eccentricities
exceeding a few $H/r$ for both disc and protoplanet
the tidal interaction may differ significantly from
the  circular disc and small protoplanet eccentricity case.
We now consider applications to the normal modes
calculated in this paper.

The equilibrium protoplanet eccentricities associated with
the two highest frequency modes
calculated in the  case
of disc model A with  two interior protoplanets
are shown in  figure \ref{fig1} while those
corresponding to disc model $B$ are illustrated in figure
\ref{fig2}. These modes  are associated with
significant internal protoplanet eccentricities
and can be thought of as giving the disc response to external
forcing. Equilibrium eccentricities corresponding to the normal
modes are also plotted in figures \ref{fig1} and \ref{fig2}.
These are generally larger than the disc eccentricities
in the inner parts of the disc.

Equilibrium protoplanet eccentricities associated with 
the modes  calculated in the  case 
of disc model A with one interior protoplanet
are shown in  figure \ref{fig5} while those
corresponding to disc model $B$ are illustrated in figure
\ref{fig5b}. Equilibrium eccentricities in the case of isolated disc model
$A$ are plotted in figure \ref{fig6} while those corresponding
to disc model $B$ are given in figure \ref{fig6b}.
In all of these cases the form of the equilibrium eccentricity
curve  tracks that of the corresponding normal mode according to
increasing number of nodes. Thus the mode with the largest number of nodes
has associated equilibrium eccentricity with the largest number of nodes.

By comparing the equilibrium eccentricities with their corresponding normal modes
one sees that the modes with the smallest frequencies or pattern speeds
in absolute magnitude tend to have high  equilibrium eccentricities
several times larger than  the disc eccentricity. These correspond to the longest
wavelength curves in figures \ref{fig5} and \ref{fig6} with corresponding
modes plotted in figures \ref{fig3} and \ref{fig4}.
From our discussion above these are expected to facilitate high embedded
protoplanet eccentricities. The reason the protoplanet eccentricity
is significantly larger than the disc eccentricity for these modes
is that, for the disc mode the effects of the nonaxisymmetric
forces due  to self-gravity  and pressure which drive the eccentricity
(see equation (\ref{ENM})) tend to cancel. 
However, the protoplanet  is subject only to self-gravity
with no cancelling effects from pressure forces.
Therefore  the equilibrium  eccentricity is larger. 
Note  that these low frequency modes are essentially disc modes
and are associated with low interior protoplanet eccentricities
when the latter are present.  Such disc modes may also be associated
with high embedded protoplanet eccentricities through 
secular resonances ( see section \ref{s3} above).
In our models such resonances occur when the pattern speed is slightly
prograde. An example is shown in figure \ref{fig5b}. This occurs for the one
protoplanet model with disc $B$ at about $20$ times the radius of the disc inner
edge. Such a resonance also occurs when disc model $A$ is used but for a higher
order mode at smaller radii (see figure \ref{fig5}).

To give numerical examples we first consider the two protoplanet
model with disc $A$ as an approximation to the
 Upsilon  Andromedae system. Taking the situation represented in
figure \ref{fig1} with inner boundary disc eccentricity
$0.1,$  for a  semi-major axis of $2.57$~AU for the outer protoplanet
in the inner cavity, being $0.6$ of the inner boundary
radius,   $r=1.5$ corresponds to
 $6.4$~AU. For 
 the mode with $ \Omega_p = 6.99\times 10^{-4}$
 and disc model $A,$
 $\Sigma=136 gm cm^{-2}$ at $r = 1.5.$
At  this radius $e=0.07,$ and the equilibrium
eccentricity $ e_p=0.1.$ 
 From figure \ref{figMIG} the orbital migration
rate is very  small  and possibly  outwards. We also find 
 $|t_e| = 10^4 (3M_{\oplus}/m_p) $~yr and  
 $|(\omega_{pg} -\Omega_p) t_e| \sim  30.0 (M_{\oplus}/m_p).$
Thus in this case the eccentric disc
has significant effects on the tidal torques acting on an embedded 
protoplanet.

The eccentricity of the outermost protoplanet  orbit  in
the inner cavity 
would be $\sim 0.1.$ The currently observed value which is three or so
times larger would be attained if the disc inner boundary was scaled
to be at twice the outer protoplanet semi-major axis.
Very similar results for the migration and circularization rates
 to those obtained above would still apply.

As an example to illustrate a case involving a very low frequency mode
we consider the one protoplanet model with disc $A.$
The modes are plotted in figure \ref{fig3} and the
equilibrium eccentricities in figure \ref{fig5}.

For 
the mode with $ \Omega_p = 4.02\times 10^{-6},$
we find that 
at $r=10$  $e_p =2.5 e.$
Supposing that $r=10$ corresponds to $5$~AU, $\Sigma=426 gm cm^{-2}.$  
From figure \ref{figMIG} we find that for $e=0.05,$  there is
torque reversal. 
 We also again find that 
 $|t_e| \sim 3\times 10^3 (3M_{\oplus}/m_p)$~yr and  
 $|(\omega_{pg} -\Omega_p) t_e| \sim 30.0 (M_{\oplus}/m_p).$            

These examples indicate  that protoplanets embedded in eccentric discs
will be in eccentric orbits and that based on resonant torque calculations
orbital migration slows down and 
 may  even reverse to become outward when the 
protoplanet eccentricity is sufficiently large.
However, we should emphasize that
these calculations are approximate and somewhat uncertain
due to the cancellation of torques arising at inner and 
outer Lindblad resonances. To further examine the issue 
of protoplanet disc tidal interaction at a
large eccentricity ( compared to $H/r$ and $e$)
we present below a simpler calculation based on local dynamical
friction which should apply in the appropriate limit
and which is in essential qualitative  agreement with 
the resonant  torque calculations.

\subsection{Dynamical friction calculation}
We here consider the case when $e_p$ significantly exceeds $H/r$
but is still also significantly less than unity.
This can be realized in sufficiently thin discs.
In this situation the motion of the protoplanet through the disc is supersonic
so we neglect pressure forces.
In addition the scale of the response in both space and time
becomes local ( even though the protoplanet
may move globally through the disc).
 For example, for a length scale H, the response time scale
would be $\sim H/(e_pr\Omega) \ll \Omega^{-1}.$

Accordingly we work in a reference frame moving instantaneously
with the protoplanet in which the disc material appears
to move with velocity ${\bf v}.$ Adopting local Cartesian
coordinates, we suppose that the perturbing potential due to
the protoplanet may be written as Fourier integral
\begin{equation} \Phi_p'({\bf r}) =\int \Phi_p'({\bf k})\exp(i{\bf k}{\bf r}) 
d^N{\bf k}. \end{equation}
Here we shall consider both the two dimensional case $(N=2)$
and the three dimensional case $(N=3)$.
For the two dimensional case with
$\Phi_p'({\bf r}) =-Gm_p/\sqrt{r^2+b^2},$ $\Phi_p'({\bf k})
= -Gm_p\exp(-bk)/(2\pi k),$ with $k=|{\bf k}|.$
In the three dimensional case with
$\Phi_p'({\bf r}) =-Gm_p/r,$ $\Phi_p'({\bf k})
= -Gm_p/(2\pi^2 k^2).$

In each case, assuming a local steady
state,  the velocity  ${\bf v}'$ induced by the protoplanet
is found from

\begin{equation}
{\bf v}\nabla {\bf v}' = -\nabla \Phi_p' \label{motof}.
\end{equation}             
We also have 
\begin{equation}
\nabla \cdot(D {\bf v}')= -\nabla(\cdot D'{\bf v}),
\end{equation}      
where $D$ denotes the density $(N=3)$ or the surface
density $(N=2).$
In this case where only inertial terms are retained
the analysis is the same as would be performed
for collisionless particles ( eg Tremaine \& Weinberg 1984).
  In this context we note that this type of calculation is
 also applicable to the frictional  interaction with material in planetesimal
 form  as well as with  the gas disc provided the surface density and
 disc thickness are appropriately specified.

The rate of change of disc momentum  ${\bf{\dot P}},$
which gives rise to a frictional force
on the protoplanet acting in the direction
of it's relative velocity,
may then be calculated from
\begin{equation}
{\bf {\dot P}} = -\int D'\nabla \Phi_p' d^N{\bf r}
\end{equation}              
Performing an integration by parts and   working in terms of
the Fourier transforms of perturbations, one obtains
\begin{equation}
{\bf {\dot P}} \cdot {\bf v}=
 - (2\pi)^N \int D \Phi_p'({\bf k}) i{\bf k}\cdot {\bf v}'({\bf k})d^N{\bf k} 
\label{fric}.
\end{equation} 
From (\ref{motof}) one obtains
\begin{equation}  
{\bf v}'({\bf k})\cdot{\bf k} = {-k^2\Phi_p'({\bf k})\over {\bf v}\cdot{\bf k}}.
\end{equation}  
In order to perform the integral (\ref{fric}) one has to apply a Landau
prescription by adding an infinitessimally  small negative imaginary
part  to the denominator ${\bf v}\cdot{\bf k}.$
One then finds
\begin{equation}
{\bf {\dot P}} \cdot {\bf v}= -{\pi D (Gm_p)^2\over v}{\cal Q},
\label{fric1}
\end{equation}
where for $N=3,$ ${\cal Q}=4\ln(k_{max}/k_{min}),$
and  for  $N=2,$ ${\cal Q}= 1/(2b).$ 
Here $(k_{max},k_{min})$ are the usual upper and lower
wavenumber cut offs (eg. Tremaine \& Weinberg 1984).
Here reasonable values are $k_{max} = v^2/(Gm_p),$ and
$k_{min} = 1/H.$
The two and three dimensional cases are thus
of the same form.
The logarithmic factor is generally of order unity.
Comparison of the two and three dimensional cases
suggests that $\rho \ln(k_{max}/k_{min}) =\Sigma/(8b).$
Thus the adopted
softening parameter $b$ should be somewhat smaller than $H.$

Using (\ref{fric1}) we  may evaluate, remembering
that ${\bf v}$ is the relative velocity between
disc and protoplanet,  the average rate of change
of angular momentum of the protoplanet from
\begin{equation}
\left \langle {dJ\over dt}\right \rangle = {\omega\over 2\pi}
\int{\pi D (Gm_p)^2 ({\bf r}\times{\bf v})\cdot{\hat{\bf k}}\over v^3}
{\cal Q} dt
\label{fric3}, \end{equation}
with the integral being taken round the orbit
and ${\hat{\bf k}}$ being the unit vector in the direction normal
to the disc.

In the two dimensional case for $D \equiv \Sigma \propto r^{-3/2}$
one obtains for small $e_p$ and $e =0.$
\begin{equation} 
{1\over J}\left\langle{dJ\over dt}\right\rangle \sim  {\pi m_p \omega \Sigma(a_p) a_p^3
\over 2 b e_p M_*^2} \label{fric4}.\end{equation}
As this is positive it corresponds to {\it outward}
migration as long as the eccentricity can be maintained.
This occurs because the disc flow tends to speed
up the protoplanet at apocentre where most
time is spent.
Numerically for $b=H/\sqrt{2}$
\begin{equation}
{t_{J}\over N_t} = - 2.6\times 10^6\left({ H\over 0.07 a_p}\right)^2 
\left({ e_p a_p\over H}\right)  {\rm yr}.
\end{equation}

While this agrees in form and is of comparable magnitude
to what  is obtained from resonant torque calculations
(see figure \ref{figMIG})
it is impractical to perform the latter at higher eccentricities
where better agreement might be attained
because of the small scale of the interaction
(even at  the  eccentricities plotted, over $10^6$ resonances
were included).

 We may also use the above formalism to calculate
the mean rate of change of orbital energy for the protoplanet
to be
\begin{equation}
\left \langle {dE\over dt}\right \rangle = -{\omega\over 2\pi}
\int{\pi D (Gm_p)^2 {\bf v}\cdot ({\bf v} -
GM_* r^{-1/2} {\hat {\mbox{\boldmath$\varphi$}}})\over v^3}
{\cal Q} dt
\label{fric3E}, \end{equation}

\noindent with $E = - {GM_* m_p\over 2 a_p}$ and
${\hat {\mbox{\boldmath$\varphi$}}}$ being the unit vector 
in the azimuthal direction.

\noindent Performing the integration we find  
for small $e_p$ and assuming $\Sigma \propto r^{-n}$ that
\begin{equation}
{1\over E}\left\langle{dE\over dt}\right\rangle 
\sim  { m_p \omega \Sigma(a_p
) a_p^3
\over 2 b e_p M_*^2} 
\left(4.16 - 7.17n \right)\label{fric5}.\end{equation}          

From this we see that, for this circular disc case,
the mean rate of change of orbital energy increases for $n < 0.6.$
This simplified calculation
indicates outward migration for density profiles
that do not increase too rapidly inwards in line with the idea
that the effect is caused by the interaction at apocentre.

 We  further comment that the more sensitive dependence on the softening
parameter means that two dimensional calculations of the type 
carried out here, require
a precise specification of this parameter
that  correctly  represents three dimensional
effects in order for them to be very accurate.
Thus two dimensional torque calculations and use
of torque formulae such as (\ref{EFF2}) suffer
from a number of uncertainties which can be of comparable
importance.

\section{Discussion } \label{conclusion}
In this paper we have calculated 
global $m=1$ modes with low
pattern speed corresponding
to introducing a finite disc   eccentricity.
An important aspect was the inclusion of self-gravity
which is important for the structure of these
modes as well as determining the motion
of embedded protoplanets.

We  considered disc models that were isolated
or contained one or two
protoplanets orbiting in an inner cavity.
In all cases global modes were found that could
be global on scales up to one hundred times the inner
cavity radius. The modes could be considered as being of
a type that were strongly coupled to the inner protoplanets
or essentially free disc modes. In the former case the disc eccentricity
could be comparable to that of the protoplanets for up to three
times the outer protoplanet orbital semi-major axis with
apsidal line antialigned with that of the protoplanet orbits. 
In the latter case the inner protoplanet orbital eccentricities
were small compared to that found  in the disc.

We  went on to  discuss the
motion of a protoplanet  embedded in an eccentric disc  
and determined, initially neglecting
tidal torques, the equilibrium (non precessing)  orbits  which maintain
apsidal alignment with the disc gas orbits.
Equilibrium eccentricities were found to be comparable or possibly even exceed
the disc eccentricity. In some cases secular resonance could occur
producing particularly large protoplanet eccentricities.

 We then formulated the calculation of the
 response of an eccentric disc to a  protoplanet 
in the earth mass range
 in order to determine the time rate of change of the eccentricity and
 orbital migration rate. We found that equilibrium
 aligned orbits with very similar
 eccentricity to that of the gas disc may suffer no eccentricity change
 while undergoing inward migration in general. 
This was found from the resonant torque calculations
but  is also expected from direct consideration of the equations
governing the tidal response. 
However, when the non precessing equilibrium aligned orbit has a
significantly higher eccentricity than the disc, as can
occur generally, but in particular
for modes with very small pattern speed, orbital migration
may be significantly reduced or reverse from inwards to outwards
for the disc models we considered.

 Attainment of high eccentricities
 in this way typically requires the characteristic test particle orbit
precession frequency or mode pattern speed to significantly
exceed the characteristic orbital circularization rate,
a situation more likely for lower mass protoplanets.
When tidal circularization dominates, the protoplanet
equilibrium eccentricity
is reduced while the  apsidal line becomes significantly
inclined to that of the disc.
 However, high protoplanet eccentricities  could be  excited
by  gravitational interactions between them (Papaloizou \& Larwood 2000)
and under favourable conditions
this effect could act to counter tidal circularization
generating  significantly
higher protoplanet  eccentricities  than
that of the disc. This will be  a topic for future investigation.

Although there is some uncertainty in the resonant torque
calculations because of the need to sum contributions
of different sign,
weakening of the tidal interaction is expected on general
physical grounds on account of larger protoplanet disc 
relative velocities at higher eccentricity. 
This indication  of migration reversal
at the higher eccentricities
was found to be  supported by a local dynamical friction calculation
applicable in that limit.
In this case the interaction near apocentre tends to speed
up the protoplanet while the interaction near pericentre tends
to slow it down. These effects are of opposite sign
but the longer time spent near apocentre results in a
net outward migration of the protoplanet for the surface
density considered.

Thus the existence of global non circular  motions
in discs with radial excursions comparable to or exceeding  the

\noindent semi-thickness
may have important consequences for the migration
of cores in the earth mass range. While processes of the type considered
in this paper are unlikely to lead to the very high eccentricities
observed for some giant planets, they may be important in controlling
migration during planet formation as well as
producing modest eccentricities $\sim 0.2.$

\begin{acknowledgements}
The author thanks the IAP for visitor
support and caroline Terquem for valuable
and stimulating discussions as  well as a carefull reading
of a preliminary draft of this paper.
\end{acknowledgements}


\vspace{20mm}

\begin{appendix}\label{A1}

\title{{\bf APPENDIX 1}}
\section*{The time averaged potential  due to a perturbing inner planet}
\noindent The gravitational potential per unit mass
due to a planet of mass $m_p$
located at $ {\bf r_p} \equiv (r_p,\varphi_p)$
at $ {\bf r} \equiv (r,\varphi)$ is
\be \Phi = - {Gm_p\over \sqrt{r^2+r_p^2 -2rr_p\cos(\theta)}},\ee
with $\theta =\varphi - \varphi_p.$

\noindent In order to incorporate the effects of protoplanets orbiting
interior to the disc  we adopt a Jacobi coordinate system.
In this system the coordinates of the innermost protoplanet
are referred to the central star. The coordinates of the remainder are referred
to the centre of gravity of the central mass and all interior
protoplanets. The disc is referred to the centre of mass of central star
and all inner protoplanets.

This has the following consequences:

\noindent For an object interior to the                                   
protoplanet with $r < r_p,$ the acceleration of the coordinate system
due to the protoplanet must be allowed for. This gives rise,
correct to first order in $m_p,$ to the indirect
potential ( eg. Brouwer \& Clemence 1961)
\be \Phi_i =  {Gm_p r\cos(\theta)\over r_p^2 },\ee
to be added to the potential $\Phi.$

\noindent For an object exterior to the protoplanet one must take account
of the fact that the coordinate system is now based on the centre of mass
of the inner protoplanets and central star. Assuming initially
that $m_p$ is the only such protoplanet, the central potential
is modified to become

\be \Phi = - {GM_* \over {|{\bf r} + m_p {\bf r_p}/M_*| }}.\ee
To first order in $m_p$ this gives
\be \Phi = - {GM_* \over r}     +  {Gm_p r_p \cos\theta \over r^2}   .\ee
In this case too the additional potential on the right hand side
of the above can be incorporated
into the perturbing planet potential and  viewed as giving
rise to an indirect potential. Thus the form of the perturbing
potential

\be \Phi = - {Gm_p\over \sqrt{r^2+r_p^2 -2rr_p\cos(\theta)}}
+ {Gm_p rr_p \cos\theta \over max(r^3,r_p^3)},\ee                         

\noindent incorporates both cases $r < r_p$ amd $r_p <r.$

\noindent In addition, although we included just one inner perturbing
protoplanet, because we work to first order in their masses,
the principle of linear superposition is valid such that
the contributions of many such objects may be linearly superposed.

The perturbing
potential due to a single
protoplanet may be decomposed as a Fourier expansion in $\theta.$

Thus

\be {\Phi \over Gm_p}
= -\sum_{m=0}^{\infty} { K_m(r,r_p) \over \pi(1+ \delta_{m0})}
\cos m \theta ,\ee

with

\be K_m(r,r_p) = -{1\over G m_p}
\int^{2\pi}_0 \Phi \cos m\theta d\theta .\ee

For the problem on hand, namely the study of global $m=1$ modes
of the disc planet system, we need only consider $m=0$ and $ m=1.$
Then from the above
\be {\pi\Phi \over G m_p} =
-{1\over 2}K_0(r,r_p) - K_1(r,r_p)\cos \theta ,\ee

with

\be K_0(r,r_p) = 
\int^{2\pi}_0 {1\over \sqrt{r^2+r_p^2 -2rr_p\cos(\theta)}}d\theta \ee

and

 $$ K_1(r,r_p) = \left(
\int^{2\pi}_0
{1\over \sqrt{r^2+r_p^2 -2rr_p\cos(\theta)}}\cos\theta d\theta\right.  $$
\be  \hspace{30mm}   \left.  - { \pi rr_p \over max(r^3,r_p^3)}\right).\ee

\section*{ Time averaged potential for a protoplanet with small eccentricity}

We now suppose the protoplanet has a small eccentricity, $e_p$
and write for its motion
$r_p = a_p(1+e_p\cos\omega t),
\varphi_p = \omega t -2e_p\sin\omega t,$ where without loss of generality
we assume the apsidal line to be along $\varphi =0$
where the displacement $\xi_r(r_p) = e_p a_p$ at $t=0.$

\noindent Here the protoplanet
semi-major axis is $a_p$ and the orbital frequency
is $\omega.$

\noindent Expanding to first order in $e_p,$ we find for the
single protoplanet perturbing potential

$$ {\pi \Phi \over Gm_p} = -{1\over 2}K_0(r,a_p) - 
{a_p e_p\over 2} {\partial K_0(r,a_p)\over \partial a_p}\cos\omega t - $$
$$ \left(K_1(r,a_p) + a_p e_p {\partial K_1(r,a_p)\over \partial a_p}
\cos\omega t\right)
\cos(\varphi -\omega t +2e_p\sin\omega t)  .$$

\noindent Performing the time average then gives
\be {\Phi \over Gm_p} = -{1\over 2\pi}K_0(r,a_p)
 -{e_p\over 2a_p \pi} {\partial ( a_p^2 K_1(r,a_p)) \over \partial a_p}
\cos\varphi .\ee

\noindent For small $e_p$ we may replace $a_p$ by $r_p$ and
expressing the result in terms of  the radial
displacement $\xi_r(r_p),$ we find

\be {\Phi \over Gm_p} = -{1\over 2\pi}K_0(r,r_p)
-{\xi_r(r_p)\over 2r_p^2\pi} {\partial (r_p^2 K_1(r,r_p))\over \partial r_p}
\cos\varphi . \label{appot}\ee

\noindent From equation (\ref{appot}) the time averaged
potential due to a protoplanet in circular orbit $(\xi_r(r_p) = 0 )$
is
\be \Phi_{ext} =  -{Gm_p\over 2\pi}K_0(r,r_p).\ee

\noindent When $\xi_r(r_p) \ne 0,$ equation  (\ref{appot})
gives the $m=1$ component of the perturbing potential as the real part of
$\Phi_p'(r)\exp(i\varphi),$ with 

 \be \Phi_p'(r) = -{G m_p \xi_r(r_p)\over 2\pi r_p^2}
{\partial (r_p^2 K_1(r,r_p))\over \partial r_p}. \ee

\end{appendix}


\begin{appendix}\label{A22}

\title{{\bf APPENDIX 2}}
\section*{The equations of motion in two dimensional orthogonal cordinates}
The basic equations of motion  (\ref{12d})  written in vector form are
\begin{equation}  {\partial  {\bf v} \over\partial t}+
{\bf v}\cdot \nabla {\bf v} =
-{1\over \Sigma}\nabla \Pi -
\nabla \Phi.
\label{1A2d} \end{equation}

We wish to write these in component form in a general two dimensional
orthogonal coordinate system $(a,\lambda)$ in the Cartesian $(x,y)$
plane as introduced 
in section (\ref{s4}). The unit vectors in these orthogonal
coordinate directions
are ${\bf i}_a = \nabla a/|\nabla a|$ and  $  {\bf i}_{\lambda}
= \nabla \lambda/|\nabla \lambda|$
respectively.    
The orthogonal vertical coordinate $z$ has an additional  associated
orthogonal unit vector
${\bf {\hat k}} ={\bf i}_a \times {\bf i}_{\lambda}.$
However, there is no dependence on $z$ and the velocity component
in that direction  can be taken to be zero.

\noindent Thus we may write
${\bf v}\equiv  ( v_{a}, v_{\lambda})$
or \be {\bf v} =  v_{a} {\bf i}_a + v_{\lambda} {\bf i}_{\lambda}. \ee

\noindent To deal with equation (\ref{1A2d}), we first use the vector identity
\be {\bf v}\cdot \nabla {\bf v}= {1\over 2} \nabla( |{\bf v}|^2)
+   {\mbox{\boldmath $\omega$}}   \times {\bf v} ,\ee
where $   {\mbox{\boldmath $\omega$}}
= \nabla \times {\bf v} $ is the vorticity (Arfken \& Weber  2000).

Because  we only wish to consider the equations in a two dimensional
limit , only the $z$ component of vorticity, $\omega_z$  is non zero
so that we may write equation (\ref{1A2d}) as

\begin{equation}  {\partial  {\bf v} \over\partial t}+
 \omega_z {\bf {\hat k} }\times {\bf v} +{1\over 2} \nabla( |{\bf v}|^2)=
-{1\over \Sigma}\nabla \Pi -
\nabla \Phi.
\label{1A22d} \end{equation}         

We may now write equation (\ref{1A22d} ) in component form
using the identities (see Arfken \& Weber 2000)

\be \nabla \equiv \nabla a {\partial \over \partial a}
+ \nabla \lambda  {\partial \over \partial \lambda } \ee  
and

\be 
\omega_z =\left(\nabla \times {\bf v}\right) \cdot {\bf {\hat k} }
=|\nabla \lambda||\nabla a|
\left ({\partial ( {v_{\lambda} \over |\nabla \lambda|})\over \partial a} 
- {\partial  ({ v_{a} \over |\nabla a|})
\over \partial \lambda } \right ). \label{VORT}\ee 

Doing this we obtain
\begin{equation}  {\partial   v_a \over\partial t}-
\omega_z  v _{\lambda} +{1\over 2} |\nabla a| {\partial 
( v_a^2 + v_{\lambda}^2)  \over \partial a} =
-{1\over \Sigma}|\nabla a| {\partial  \Pi \over \partial a}  -
|\nabla a| {\partial  \Phi \over \partial a}.
\label{1A23d} \end{equation}             

\noindent and

\begin{equation}  {\partial  v_{\lambda} \over\partial t}+
\omega_z  v _{a} +{1\over 2} |\nabla \lambda | {\partial
( v_a^2 + v_{\lambda}^2)  \over \partial \lambda } =
-{1\over \Sigma}|\nabla \lambda | {\partial  \Pi \over \partial \lambda}  -
|\nabla \lambda | {\partial  \Phi \over \partial \lambda}.
\label{1A24d} \end{equation}       

After inserting (\ref{VORT}) into (\ref{1A23d}) and (\ref{1A24d})
it is a simple matter  to obtain  
equations (\ref{12d1}) and (\ref{12d2}) as given in section (\ref{s4}).

\end{appendix}

\end{document}